\documentclass[twocolumn,preprintnumbers,secnumarabic,amsmath,amssymb,groupedaddress,nofootinbib]{revtex4-2}
\usepackage{mathrsfs}
\usepackage{amsmath, amsthm, amssymb}
\usepackage{color}
\usepackage{epstopdf}
\usepackage[pdftex]{graphicx}
\usepackage{verbatim}
\usepackage{hyperref}
\usepackage{enumitem}
\usepackage[normalem]{ulem}
\usepackage{accents}
\usepackage{tikz}
\usetikzlibrary{decorations.markings}
\usetikzlibrary{decorations.pathmorphing}
\tikzset{snake it/.style={decorate, decoration=snake}}

\makeatletter
\def\cstar#1{\expandafter\@cstar\csname c@#1\endcsname}
\def\@cstar#1{\ifcase#1\or $^*$\or $^*$\or $^*$\fi}
\AddEnumerateCounter{\cstar}{\@cstar}{$^*$}
\makeatother

\begin{document}

\title{PQ Axiverse}

\author{Mehmet Demirtas$^{a,b}$}

\author{Naomi Gendler$^{c}$}

\author{Cody Long$^{d}$}

\author{Liam McAllister$^{c}$}

\author{Jakob Moritz$^{c}$}

\affiliation{${}^a$ The NSF AI Institute for Artificial Intelligence and Fundamental Interactions}
\affiliation{${}^b$ Department of Physics, Northeastern University, Boston, MA 02115, USA}
\affiliation{${}^c$ Department of Physics, Cornell University, Ithaca, NY 14853, USA}
\affiliation{${}^d$ Department of Physics and CMSA, Harvard University, Cambridge, MA 02138, USA}

\begin{abstract}
We show that the strong CP problem is solved in a large class of compactifications of string theory.  The Peccei-Quinn mechanism solves the strong CP problem if the CP-breaking effects of the ultraviolet completion of gravity and of QCD are small compared to the CP-preserving axion potential generated by low-energy QCD instantons.  We characterize both classes of effects.
To understand quantum gravitational effects, we consider an ensemble of flux compactifications of type IIB string theory on orientifolds of Calabi-Yau hypersurfaces in the geometric regime, taking a simple model of QCD on D7-branes.  We
show that the
D-brane instanton
contribution to the neutron electric dipole moment falls exponentially in $N^4$, with $N$ the number of axions. In particular,
this contribution is negligible in all models in our ensemble with $N>17$.  We interpret this result as a consequence of large $N$ effects in the geometry that create hierarchies in instanton actions and also suppress the ultraviolet cutoff.
We also compute the CP breaking due to high-energy instantons in QCD. In the absence of vectorlike pairs, we find contributions to the neutron electric dipole moment that are not excluded, but that could be accessible to future experiments if the scale of supersymmetry breaking is sufficiently low.
The existence of vectorlike pairs can lead to a larger dipole moment.
Finally, we show that a significant fraction of models are allowed by standard cosmological and astrophysical constraints.
\end{abstract}

\maketitle

\section{Introduction}
\label{sec:intro}

Experiments reveal that QCD preserves CP symmetry to very high precision.  The action for QCD can in principle include the CP-violating term
\begin{align}\label{eq:CSterm}
\mathcal{L} \supset
\frac{\theta}{32\pi^2} \epsilon^{\mu \nu \rho \sigma} \text{tr}\, G_{\mu \nu} G_{\rho \sigma}\, ,
\end{align}
in conventions where the quark mass matrix is real. CP is preserved by QCD if $\theta=0$, and
experiments show that $\theta \lesssim 10^{-10}$ in our universe \cite{nEDM:2020crw,PhysRevLett.97.131801,PhysRevD.92.092003}.
The strong CP problem is the question of why this number is so small.

There are several possible solutions to this problem, but arguably the most viable proposal, and the one that we will consider in this work, is the Peccei-Quinn (PQ) mechanism \cite{Peccei:1977hh}.
In this scenario, the QCD $\theta$-angle is promoted to a dynamical pseudoscalar $\theta(x)$, the \textit{QCD axion}, with a global shift symmetry $\theta(x)\mapsto \theta(x)+const$ that is broken by QCD instantons.

The potential induced by low-energy QCD instantons is (see e.g.~\cite{GrillidiCortona:2015jxo}) \begin{align}\label{eq:vqcd}
V_{\mathrm{QCD}}(\theta) = \frac{1}{2}\Lambda_{\mathrm{QCD}}^4 \frac{m_u m_d}{(m_u+m_d)^2} \theta^2 + \mathcal{O}\bigl(\theta^4\bigr)\, ,
\end{align} where $ \Lambda_{\mathrm{QCD}}^4 = f_\pi^2 m_\pi^2$, $f_\pi \approx 92~\mathrm{MeV}$ and $m_\pi = 135~\mathrm{MeV}$ are the pion decay constant and mass, and $m_u$ and $m_d$ are the $u$ and $d$ quark masses.
Provided that \eqref{eq:vqcd} is the only significant potential term for $\theta$, minimizing the potential gives $\langle \theta\rangle=0$, and the strong CP problem is solved: the physical QCD $\theta$-angle $\theta$ is zero, in accord with experiments.

The \emph{Peccei-Quinn quality problem} is the fact that sufficiently large breaking of the QCD axion shift symmetry in another sector can perturb the minimum away from  $\theta=0$.  Without loss of generality, the potential for $\theta$ given as a linear combination of $N$ pseudoscalar fields $\xi^A$, $A=1,\ldots,N$ can be written as
\begin{align}\label{eq:vqcdpluscorrection}
V(\xi^A) = V_{\mathrm{QCD}}(\theta) + V_{\mathrm{hidden}}(\xi^A)\,,
\end{align} with $V_{\mathrm{QCD}}(\theta)$ given by \eqref{eq:vqcd}.  If the hidden sector potential $V_{\mathrm{hidden}}(\xi^A)$ is such that the minimum of $V(\xi^A)$ occurs at $|\langle \theta\rangle| \gtrsim   10^{-10}$, we say that the PQ solution of the strong CP problem is \emph{spoiled}.
The goal of this work is to assess the PQ mechanism in string theory by calculating $V_{\mathrm{hidden}}$ in a large class of string compactifications.

The first part of this endeavor concerns the ultraviolet (UV) completion of QCD.
We note that $\theta$ can receive a contribution from QCD instantons whose zero modes are lifted by higher-dimension fermionic operators, for example four-quark operators.
We compute these effects, making three assumptions: the Standard Model is completed into a supersymmetric theory at some high scale $M_{\mathrm{SUSY}} > 1\,\mathrm{TeV}$; the charged spectrum of QCD is given by Standard Model quarks and their superpartners up to a sufficiently high scale (though we will also comment on the inclusion of vectorlike pairs); and CP is not an approximate symmetry of the ultraviolet theory.
We find that although high-scale QCD instantons do not spoil the Peccei-Quinn solution,
for sufficiently low supersymmetry breaking their contribution to $\theta$ is larger than that of CP-violating effects in the weak sector of the Standard Model.  In future experiments on the neutron electric dipole moment, such effects might be detectable.

In the second part of this work, we turn to effects of the UV completion in quantum gravity.
Preserving the success of the PQ mechanism amounts to requiring that the breaking of the axion shift symmetry by the hidden sector potential $V_{\text{hidden}}(\xi^A)$ is at least $10^{-10}$ times smaller than the breaking by low-energy QCD.
This means that simply imposing a weakly broken global shift symmetry onto the effective field theory cannot by itself solve the strong CP problem. In particular, even Planck-suppressed operators can spoil the PQ mechanism, and so a full resolution of the problem requires information about quantum gravity
contributions to axion potentials \cite{Georgi:1981pu,Holman:1992us,PhysRevD.46.539,Kamionkowski:1992mf}.
In many corners of string theory, axions --- i.e.~pseudoscalar fields with approximate shift symmetries --- are ubiquitous, and in many constructions the existence of an axion that couples to QCD is automatic.
Moreover, in vacua of string theory that furnish weakly-coupled UV completions of gravity, such as weakly-curved compactifications with small values of the string coupling, one can --- and we will --- compute the leading quantum gravity contributions to the axion potential.
String theory thus provides a natural framework
for assessing the quality problem.

We study the Peccei-Quinn quality problem in a well-understood class of solutions of string theory: compactifications of type IIB string theory on O3/O7 orientifolds of Calabi-Yau threefold hypersurfaces in toric varieties.
In this setting, we suppose that QCD is realized on a stack of D7-branes wrapping a four-cycle $D_{\mathrm{QCD}}$, so that
the QCD axion $\theta$ arises from reduction of the Ramond-Ramond four-form $C_4$ on $D_{\mathrm{QCD}}$.  The issue is that $\theta$ obtains a potential not only from QCD instanton effects, but also from \emph{stringy instantons}: that is, Euclidean D3-branes (or D7-brane gaugino condensates) wrapping various other four-cycles in the manifold.
Thus, the quality problem manifests itself through \emph{D-brane instantons that generate the hidden-sector potential for the axions $\xi^A$, $A=1,\ldots N$}.
Solving the quality problem in this context then amounts to
ensuring that these instanton effects
have negligible impact on the minimum of the QCD axion potential.

The sizes of the instanton couplings in a given compactification
can be obtained, up to a few caveats that we will discuss, by computing the volumes of four-cycles in the internal space: in particular, instantons wrapping large cycles give small couplings.
In this work, by computing the volumes of four-cycles in an ensemble of more than 100{,}000 compactifications, we
obtain upper bounds on the CP-breaking couplings encoded in $V_{\text{hidden}}(\xi^A)$. With this in hand, we compute the value of $\theta$ for each model.\footnote{The data of a model includes specification of a Calabi-Yau compactification at a given point in its moduli space, but our results turn out to be rather insensitive to the choice of point (see \S\ref{sec:generating_models}), and thus independent of the interplay with moduli stabilization discussed e.g. in \cite{Conlon:2006tq,plauschinn,Grimm:2011dj,Broeckel:2021dpz}.}

We find that the D-brane instanton contribution to $\langle \theta \rangle$, and hence to the neutron electric dipole moment, falls approximately as $\mathrm{exp}(-c N^4)$ with $c \approx 1.8$.
In
99.7\% of all the models in our ensemble, and in particular in every model with $N>17$,
\emph{D-brane instantons do not spoil the PQ mechanism.}

Whether the high-energy QCD instantons discussed above can themselves spoil the PQ mechanism, even when D-brane instantons do not, depends on the spectrum of matter charged under QCD.  Counting vectorlike pairs is an often-subtle problem (see e.g.~\cite{Bies:2021nje}) that is beyond the scope of this work, so we make conditional statements in terms of the number $n$ of such pairs.  For $n=0$, high-energy QCD instantons are harmless, and the PQ mechanism solves the strong CP problem
whenever D-brane instanton contributions can be neglected --- which is almost always the case in our ensemble.  For $n>0$ vectorlike pairs of mass $M_V$, high-energy QCD instantons make larger contributions to the neutron electric dipole moment, and the strong CP problem is solved for some but not all values of $n$ and $M_V$.

The same geometric data needed to characterize the PQ mechanism also provide information about the relic abundances of the various species of axion dark matter, and about the couplings of light axions to photons.
Using a simplified model for the dark matter relic abundances and axion-photon couplings, we analyze the astrophysical and cosmological constraints on the axions in our ensemble.
We find that for theories in this landscape with $26 \le N \le 433$, at least $50\%$ of models are allowed both by astrophysical constraints and by bounds on the axion dark matter relic abundance.

The structure of this paper is as follows.
In \S\ref{qcdinstantons}, we explain the Peccei-Quinn quality problem in general terms.
In \S\ref{sec:qcd}, we compute the contribution of high-scale QCD instantons to the QCD axion potential in field theory. We show that these effects do not spoil the Peccei-Quinn mechanism in the absence of vectorlike pairs.
In \S\ref{sec:quality}, we describe the incarnation of the Peccei-Quinn quality problem in type IIB compactifications in which the PQ axion descends from the Ramond-Ramond four-form.
We summarize the data that one needs to obtain from string theory in order to determine whether the strong CP problem is solved via the PQ mechanism.
We then give more details in the specific context of orientifolds of Calabi-Yau threefold hypersurfaces in toric varieties, and describe how we construct an ensemble of axion effective theories. In \S\ref{sec:results}, we present the results from a large ensemble of compactifications of type IIB string theory on orientifolds of Calabi-Yau threefold hypersurfaces.  We argue that the strong CP problem is generically solved in this class of theories.  We also
identify the range of parameter space in our ensemble that is allowed by cosmological and astrophysical bounds.
We conclude in \S\ref{sec:conclusions}.
In Appendix \ref{app:masses_fs}, we detail a fast algorithm to compute masses and decay constants in our models.
In Appendix \ref{app:subleading}, we describe two classes of contributions to the scalar potential that we neglect in the main analysis, and we justify the omission of these terms.
\newpage

\section{The Peccei-Quinn quality problem}\label{qcdinstantons}

We consider a set of $N$ axions $\xi^A$, $A=1,\ldots,N$, with the QCD axion a particular linear combination thereof, $\theta:=q_A \xi^A$. The scalar potential can generally be written as
\begin{align}\label{vqcdpluscorrection}
V(\theta,\xi^A) = V_{\mathrm{QCD}}(\theta) + V_{\mathrm{hidden}}(\xi^A)\, ,
\end{align}
where $V_{\mathrm{hidden}}$ encodes all contributions to the scalar potential \textit{except} the infrared (IR) contribution from QCD encoded in $V_{\mathrm{QCD}}(\theta)$.

One can think of $V_{\mathrm{hidden}}$ as a set of operators that break the continuous shift symmetries of the axions.
To illustrate this, we temporarily suppose, as originally envisioned in \cite{Peccei:1977hh}, that the axions $\xi^A$ are the $U(1)$ phases of complex fields $\Phi^A$ that obtain non-zero vevs $\Phi^A=f^A e^{i\xi^A}$ from a symmetry breaking potential $V_{symm}(\Phi^A,\bar{\Phi}^A)\equiv V_{symm}(|\Phi^A|^2)$ that is independent of the axions.  The axions receive a potential e.g.~from  higher-dimension shift-symmetry breaking operators. In the simplest scenario of a single axion $\theta$, its shift symmetry is broken by operators such as
\begin{align}
V_{\mathrm{hidden}} \supset \sum_{n} c_{n} \frac{\Phi^{n}}{M_\mathrm{pl}^{n-4}} + c.c.
\label{eq:Vhidden1}
\end{align}
Here, without further assumptions, the contributions to the scalar potential from higher-dimension operators scale polynomially in $f/M_\mathrm{pl}$. If we take $f \sim 10^{10} \ \text{GeV}$ in order to comply with standard cosmological bounds \cite{Preskill:1982cy, Abbott:1982af, Dine:1982ah}, then we see that the operators  in \eqref{eq:Vhidden1} must be suppressed up to and including $n=10$ (we will explain more precisely exactly how much suppression is needed in what follows). For larger decay constants, even higher order operators must be suppressed. From this point of view, the strong CP problem gets traded for an (arguably just as bad) Peccei-Quinn quality problem.

In grappling with the PQ quality problem, one ought to contend with two possibilities: first, that the UV-sensitive shift symmetry breaking from \textit{small} (i.e. high-energy) gauge instantons in QCD is too large; and second, that other sources of shift symmetry breaking, unrelated to the QCD gauge instanton, perturb the minimum of the low-energy potential \eqref{eq:vqcd}.

The issue of higher-order shift symmetry breaking operators is abated, though not eradicated, in string theory \cite{Dine:1986bg}. This is because in certain well-studied corners of the string landscape, the axion shift symmetries are exact in perturbation theory, and are broken only by non-perturbative effects.  In such theories, coefficients such as $c_n$ in
\eqref{eq:Vhidden1}
are exponentially suppressed, in terms of instanton actions that are large at weak coupling.
One might hope that in some cases, this suppression
will suffice to allow the PQ mechanism to survive.
To see if this hope is realized, we will compute these coefficients in a large ensemble of string compactifications and explicitly check whether the shift symmetry is in fact preserved to the high degree necessary.

To set the stage, we first consider a four-dimensional effective field theory with a large UV cutoff $M$,
containing in particular $N$ axion degrees of freedom $\xi^A$, and a UV version of the Standard Model, but with all stringy degrees of freedom integrated out. As such, the UV Lagrangian will contain an explicit scalar potential $V_{\text{hidden}}(\xi^A)$ for the axions $\xi^A$ generated by physics at scales higher than $M$. As mentioned above, a key feature of the theories considered in this paper is that the axion shift symmetries are only broken by instanton effects, which preserve a discrete shift symmetry $\xi^A\rightarrow \xi^A+2\pi k^A$, with $\vec{k}\in \mathbb{Z}^N$. An instanton is naturally labeled by a \textit{charge} $\vec{q}\in \mathbb{Z}^N$, and by a Euclidean action $S_E^{\vec{q}}>0$, and in terms of these the scalar potential generated by UV instantons is written as
\begin{align}\label{eq:UVpotential}
V_{\text{hidden}}(\xi^A)=&M^4\sum_{\vec{q}\in \mathbb{Z}^N}A_{\vec{q}}\,e^{-S_E^{\vec{q}}}\Bigl(1-\cos(\vec{q}\cdot \vec{\xi}-\varphi_{\vec{q}})\Bigr)\nonumber\\
\equiv & \sum_{\vec{q}\in \mathbb{Z}^N}\,\Lambda^4_{\vec{q}}\,\Bigl(1-\cos(\vec{q}\cdot \vec{\xi}-\varphi_{\vec{q}})\Bigr)\, ,
\end{align}
with one-loop determinants $A_{\vec{q}}>0$, and UV CP-breaking phases $\varphi_{\vec{q}}$. Thus we see that any discussion of the PQ quality problem must contend with the possibility that the QCD axion couples to hidden sector axions in a way that shifts the minimum outside of the acceptable range. Without a theory that relates the phases $\varphi_{\vec{q}}$ in such a way that all of the terms in \eqref{eq:UVpotential} are harmless, these contributions of UV instantons pose a serious threat to the viability of the PQ mechanism.\footnote{A first-principles calculation of the $\varphi_{\vec{q}}$ is beyond the scope of this work, but the phase associated to the QCD-generated potential arises from the phases in the quark masses, and we find it implausible that such a quantity could directly relate to the phases of (all) the high-scale stringy instantons we will be considering.}

Of particular importance in \eqref{eq:UVpotential} is the charge of the QCD instanton, denoted $\vec{q}_{Q}$, which defines the QCD $\theta$-angle via the relation $\theta_Q:=\vec{q}_{Q}\cdot \vec{\xi}$. Crucially, in order for the PQ mechanism to solve the strong CP problem, the UV potential $V_{\text{hidden}}(\xi^A)$ must leave at least one linear combination of axions sufficiently light so that the low energy QCD potential is the dominant contribution to  the scalar potential along the light direction.  In this situation the QCD $\theta$-angle will remain close to zero.

The contribution to the $\theta$-angle from instantons is easily estimated as
\begin{align}
	\Delta \theta \simeq  \frac{\Lambda_{\vec{q}_c}^4}{\Lambda_{\text{QCD}}^4}\,,
\end{align}
where $\vec{q}_c$ is the charge of the instanton with smallest action such that the linear span of $\vec{q}_c$ and all instantons more dominant than $\vec{q}_c$ contains $\vec{q}_Q$. The strong CP problem is solved if $\Delta \theta<10^{-10}$, i.e. if
\begin{equation} \label{eq:instact}
\frac{M^4}{M_\mathrm{pl}^4}A_{\vec{q}_c}\, e^{-S_E^{\vec{q}_c}}\gtrsim e^{-S_{\text{crit.}}}\, ,\quad S_{\text{crit.}}\approx 200\, .
\end{equation}
The instantons that enter \eqref{eq:UVpotential} are of two distinct types: there are high-energy QCD instantons, for which $\vec{q}$ is an integer multiple of $\vec{q}_Q$ and $S_{\vec{q}_Q}=\frac{8\pi^2}{g^2(M)}$, and there are bona fide stringy instantons. Our setup will allow us to treat both types of instantons on the same footing, and the total $\theta$-angle is the sum of the contributions from these two classes of instantons:
\begin{align}\label{eq:deltathetattot}
	\Delta \theta = \Delta \theta_{\mathrm{QCD}}+\Delta \theta_{\mathrm{stringy}}\,.
\end{align}
In the next two sections we will compute these effects in turn.

We remark that in view of \eqref{eq:instact}, there appear to be three a priori independent mechanisms to lower the scale of the potential from hidden sector instantons:
\begin{enumerate}
	\item Lowering the cutoff $M\ll M_\mathrm{pl}$ suppresses the instanton potential, and it has been argued that this effect occurs naturally in gravitational theories with many light degrees of freedom \cite{Arkani-Hamed:2005zuc,Dvali:2007hz,Dvali:2007wp}. Indeed, we will find very low UV cutoffs in controlled solutions when the number of axions is large (cf.~\cite{Demirtas:2018akl}).
	\item Raising the instanton actions $S_E^{\vec{q}}$ would suppress the axion potential parametrically.  However, scaling up instanton actions homogeneously by an overall dilatation of the compact space is incompatible with keeping the QCD gauge coupling at its observed value.
	\item The one-loop factors $A_{\vec{q}}$ are parametrically small if the SUSY breaking scale is low. However, the running of the QCD coupling is also affected by the SUSY breaking scale, and it will turn out that overall the axion potential is more suppressed when the SUSY breaking scale is \textit{high}: see \eqref{eq:theta_vs_Msusy}.
\end{enumerate}

\section{QCD instantons and $\Delta \theta$} \label{sec:qcd}
We begin by studying CP-breaking effects
resulting from QCD gauge instantons.
At first glance this might seem to be a pointless exercise, because it is well-known that in a vectorlike and CP-invariant theory, such as that defined by the renormalizable QCD Lagrangian, CP is not broken spontaneously \cite{Vafa:1983tf,Vafa:1984xg}. However, already in the (renormalizable) Standard Model, QCD inherits small violations of CP from the weak interactions \cite{Georgi:1986kr}, as we will review momentarily. Furthermore, and more importantly for our purposes, the low energy potential for the QCD axion generally contains contributions from \textit{small} instantons that are sensitive to CP breaking at very high scales $M\gg 1$ TeV, which remain completely unexplored by the LHC.\footnote{The high-energy tail of the instanton potential has been analyzed in various scenarios for beyond the Standard Model physics in the literature \cite{Holdom:1985vx,Flynn:1987rs,
Choi:1998ep,Csaki:2019vte,Gherghetta:2020keg,Kitano:2021fdl}.}

Throughout most of this paper we will make the following assumptions about physics between the weak scale and the Planck scale:
\begin{enumerate}[label=(\roman*)]
	\item The Standard Model is completed into a supersymmetric theory at a scale $M_{\mathrm{SUSY}}>1$ TeV.
	\item Except for superpartners, there are no light states charged under QCD.
	\item Above a UV threshold $M$, CP is no longer an approximate symmetry.
\end{enumerate}
We will refer to (i) and (ii) collectively as the \textit{SUSY desert assumptions}.  Assumption (iii) is broadly applicable in string theory, as emphasized particularly in \cite{Dine:1986bg}.
On general grounds, it is expected that quantum gravity breaks all global symmetries in the UV; and indeed, in string compactifications with fluxes, requiring all Wilson coefficients of higher-dimension operators to be real would amount to an infinite amount of tuning imposed on the finite set of discrete data defining a model.  Finally, we note that if (iii) were not to hold, there would be no strong CP problem in the first place.

At various stages we will relax assumption (ii)  to
\begin{enumerate}[label=(\roman*\cstar*)]
	\setcounter{enumi}{1}
	\item QCD remains asymptotically free at high energies.
\end{enumerate}
This assumption allows up to \emph{three} vectorlike pairs in the charged spectrum, in addition to the superpartners of the Standard Model quarks.

With these assumptions in hand, we can examine the  instanton-generated
axion potential.  First, let us set our notation. We will work in conventions where all Standard Model fermions are left-handed Weyl fermions. We denote by $Q^i\equiv\begin{pmatrix}
u_L^i\\
d_L^i
\end{pmatrix}$, $i=1,2,3$, the three generations of quarks in the $(3,2)_{\frac{1}{6}}$ of the Standard Model gauge group $SU(3)\times SU(2)_L \times U(1)_Y$, by $U^i\equiv (u_R^i)^c$ the up-type $SU(2)$ singlets in $(\bar{3},1)_{-\frac{2}{3}}$, by $D^i\equiv ({d_R^i})^c$ the down-type $SU(2)$ singlets in $(\bar{3},1)_{\frac{1}{3}}$, and by $H$ the Higgs scalar in $(1,2)_{\frac{1}{2}}$. For each quark generation we define an axial $U(1)_A^i$ such that all quarks of the given generation have charge one, and all others are singlets. The axial $U(1)$s are   anomalous, i.e.~under
\begin{equation}
U(1)_A^i:\quad (Q^i,D^i,U^i)\longrightarrow e^{i\alpha_i}(Q^i,D^i,U^i)\, ,
\end{equation}
we have $\theta \rightarrow \theta -2\sum_i \alpha_i$. The Yukawa couplings,
\begin{equation}
\mathcal{L}_{\text{int.}}\supset y_{ij}^dD^i H^\dagger Q^j+ y_{ij}^u U^i H Q^j+c.c.\, ,
\end{equation}
break the axial $U(1)$s perturbatively, but below the mass of the Higgs we can just interpret them as Dirac mass terms $m_{ij}^{u,d}= y_{ij}^{u,d}v$ with $v$ the Higgs vev, assign them spurious $U(1)$ charges, and treat the $U(1)$s as spurious symmetries.

\subsection{Infrared contributions}\label{sec:QCD_IR_contributions}

\begin{figure}
	\centering
	\begin{tikzpicture}
	\draw[black](0,0) circle(1.5);
	\begin{scope}[decoration={
		markings,
		mark=at position 0.5 with {\arrow{latex}}}
	]
	\draw[postaction={decorate}] (.2, 1.486) -- (2,2.5);
	\draw[postaction={decorate}]  (1, 1.118) -- (2,2.5);
	\draw[postaction={decorate}]  (.2, -1.486) -- (2,-2.5);
	\draw[postaction={decorate}]  (1, -1.118) -- (2,-2.5);
	\draw[postaction={decorate}]  (1.4, .538) -- (3,0);
	\draw[postaction={decorate}]  (1.4, -.538) -- (3,0);
	\end{scope}
	\node[align=center] at (0,0) { instanton \\ of size $1/\mu$};
	\node[] at (1,2.25) {$u_R^c$};
	\node[] at (1.8,1.7) {$u_L$};
	\node[] at (2,0.7) {$d_R^c$};
	\node[] at (2,-.7) {$d_L$};
	\node[] at (1,-2.25) {$s_R^c$};
	\node[] at (1.8,-1.7) {$s_L$};
	\node[] at (2.7,2.7) {$m_u$};
	\node[] at (3.7,0) {$m_d$};
	\node[] at (2.7,-2.7) {$m_s$};
	\node[] at (5.7,0) {$ \sim \ \mu^4 \frac{\mathrm{det}(m)}{\mu^F} e^{-\frac{8\pi}{g^2(\mu)}-i\theta}$};
	\draw (2.008,2.79) -- (2.29,2.508);
	\draw (2.29,2.79) -- (2.008,2.508);
	\draw (3.34,.1414) -- (3.06,-0.1414);
	\draw (3.06,0.1414) -- (3.34,-0.1414);
	\draw (2.15,-2.65) circle(0.2);
	\draw (2.008,-2.79) -- (2.29,-2.508);
	\draw (2.29,-2.79) -- (2.008,-2.508);
	\draw (2.15,2.65) circle(0.2);
	\draw (3.2,0) circle(0.2);	
	\node at (0,-2)[circle,fill,inner sep=0.5pt]{};
	\node at (-0.77,-1.84)[circle,fill,inner sep=0.5pt]{};
	\node at (-0.39,-1.96)[circle,fill,inner sep=0.5pt]{};
	\end{tikzpicture}
	\caption{Lifting of fermion zero modes by mass insertions.}
	\label{fig:instanton1}
\end{figure}
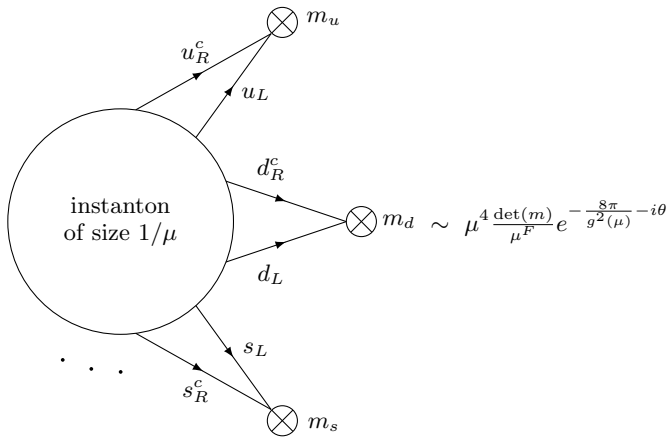

The most straightforward and well-known contribution to the instanton-generated axion potential can be written schematically as an integral
\begin{equation}\label{eq:axion_potential}
\delta V_{\text{inst.}}^{m}\sim \int \frac{d\mu}{\mu} \mu^4 \frac{\prod_{i=1}^F m_i}{\mu^{F}}e^{-\frac{8\pi^2}{g^2(\mu)}-i\theta}+c.c.\, ,
\end{equation}
where $\mu$ is the inverse size of the instanton, $F$ is the number of quark flavors with mass below $\mu$, and $m_i$ are the quark masses. The mass determinant $\prod_{i=1}^F m_i$ cancels the axial $U(1)$ charge of $e^{-i\theta}$, and equivalently lifts the fermion zero modes of the instanton background (see Figure \ref{fig:instanton1}).

Here, as usual, the contribution from large instantons in units of the QCD scale cannot actually be evaluated using the above expression, because QCD is strongly coupled in the IR. This, however, is not much of a practical issue because the IR contribution to the axion potential is determined by the pion mass and decay constant, which are known experimentally. The QCD-induced axion potential reads (see e.g.~\cite{Hook:2018dlk})
\begin{equation}\label{eq:axion_potential_QCD}
\delta V^{QCD}=-m_\pi^2 f_\pi^2\sqrt{1-\frac{4m_u m_d}{(m_u+m_d)^2}\sin\left(\frac{\theta}{2}\right)^2}\, .
\end{equation}
On general grounds, as renormalizable QCD itself does not break CP, the contribution \eqref{eq:axion_potential_QCD} to the axion potential has its minimum at $\theta=0$, leading to the PQ solution to the strong CP problem \cite{Peccei:1977hh,Vafa:1983tf,Vafa:1984xg}.

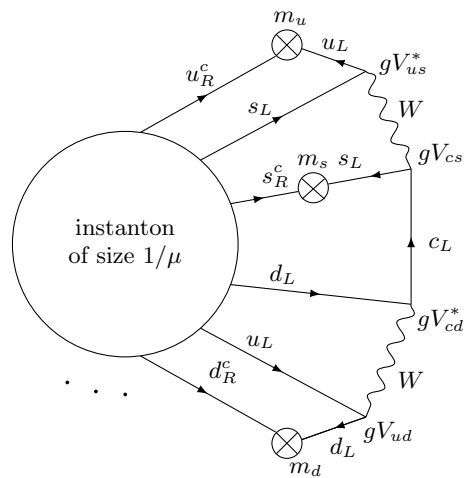
\begin{figure}
	\centering
	\begin{tikzpicture}
	\draw[black](0,0) circle(1.5);
	\begin{scope}[decoration={
		markings,
		mark=at position 0.5 with {\arrow{latex}}}
	]
	\draw[postaction={decorate}] (.2, 1.486) -- (2,2.5);
	\draw[postaction={decorate}]  (1, 1.118) -- (3.2,2.3);
	\draw[postaction={decorate}]  (.2, -1.486) -- (2,-2.5);
	\draw[postaction={decorate}]  (1, -1.118) -- (3.2,-2.3);
	\draw[postaction={decorate}]  (1.4, .538) -- (2.3,0.7);
	\draw[postaction={decorate}]  (1.4, -.538) -- (3.8,-0.8);
	\end{scope}
	\begin{scope}[decoration={
		markings,
		mark=at position 0.5 with {\arrow{latex}}}
	]
	\draw[postaction={decorate}] (3.8,1)--(2.7,.8);
	\end{scope}
	\begin{scope}[decoration={
		markings,
		mark=at position 0.5 with {\arrow{latex}}}
	]
	\draw[postaction={decorate}] (3.2,2.3) -- (2.35,2.6) ;
	\draw[postaction={decorate}]  (3.8,-.8) -- (3.8,1);
	\draw[postaction={decorate}] (3.2,-2.3) --(2.35,-2.6);
	\end{scope}
	\path [draw=black, snake it] (3.2,2.3) --(3.8,1.);
	\path [draw=black, snake it] (3.8,-.8)--(3.2,-2.3) ;
	\draw (3.2,-2.3) -- (2.35,-2.6);
	\draw (2.5,0.75) circle(0.2);
	\draw (2.35,0.87) -- (2.65,.6);
	\draw (2.35,0.6) -- (2.65,0.87);
	\node[align=center] at (0,0) { instanton \\ of size $1/\mu$};
	\node[] at (3.7,2.4) {$g V_{us}^* $};
	\node[] at (4.2,1.2) {$g V_{cs} $};
	\node[] at (4.2,-1) {$g V_{cd}^* $};
	\node[] at (3.5,-2.5) {$g V_{ud} $};
	\node[] at (2.8,2.65) {$u_L$};
	\node[] at (3.8,1.8) {$W$};
	\node[] at (3.8,-1.8) {$W$};
	\node[] at (4.2,0) {$c_L$};
	\node[] at (1,2.25) {$u_R^c$};
	\node[] at (1.8,1.8) {$s_L$};
	\node[] at (2,.9) {$s_R^c$};
	\node[] at (2.1,-.4) {$d_L$};
	\node[] at (1.3,-1.7) {$d_R^c$};
	\node[] at (1.8,-1.3) {$u_L$};
	\node[] at (3,1.1) {$s_L$};
	\node[] at (2.5,1.1) {$m_s$};
	\node[] at (2.2,3) {$m_u$};
	\node[] at (2.4,-3) {$m_d$};
	\node[] at (2.9,-2.7) {$d_L$};
	\draw (2.008,2.79) -- (2.29,2.508);
	\draw (2.29,2.79) -- (2.008,2.508);
	\draw (2.15,-2.65) circle(0.2);
	\draw (2.008,-2.79) -- (2.29,-2.508);
	\draw (2.29,-2.79) -- (2.008,-2.508);
	\draw (2.15,2.65) circle(0.2);	
	\node at (0,-2)[circle,fill,inner sep=0.5pt]{};
	\node at (-0.77,-1.84)[circle,fill,inner sep=0.5pt]{};
	\node at (-0.39,-1.96)[circle,fill,inner sep=0.5pt]{};
	\end{tikzpicture}
	\caption{Lifting of fermion zero modes by CP-breaking weak interactions.}
	\label{fig:instanton2}
\end{figure}

On the other hand, CP-breaking effects beyond QCD, e.g.~in the form of higher-dimension multi-fermion operators with complex Wilson coefficients, can also lift fermion zero modes and thus shift the minimum of the axion potential away from zero. In the Standard Model, CP breaking comes from the complex phase in the CKM matrix, and the leading correction with non-trivial CP-breaking phase to the instanton-generated axion potential is depicted in Figure \ref{fig:instanton2} \cite{Flynn:1987rs}.

Its effect is still dominated by the IR, where the instanton calculation is not reliable:
\begin{align}\label{eq:axion_potential2}
&\delta V_{\text{inst.}}^{\text{CKM}}\sim  \int \frac{d\mu}{\mu} \mu^4 \frac{\prod_{i=1}^F m_i}{\mu^{F}}\times \nonumber\\
&\mu^6\frac{G_F^2}{m_c^2}V_{ud}V^*_{cd}V_{cs}V^*_{us}\, e^{-\frac{8\pi^2}{g(\mu)^2}-i\theta }+c.c.\, ,
\end{align}
which is valid in the range $\mu<m_c$.

However, Standard Model effects induce a dimensionless CP breaking parameter in the chiral effective theory of order \cite{Georgi:1986kr}
\begin{equation}\label{eq:GRfloor}
\omega:=G_F^2 f_\pi^4 |\text{Im}(V_{ud}V_{cd}^* V_{cs}V_{us}^*)|\approx 4\times 10^{-19}\, ,
\end{equation}
so one expects $\langle \theta \rangle\sim \omega$ if $\theta$ is dynamical, and in the absence of further physics beyond the Standard Model. Thus, if there exists a QCD axion, measuring a QCD $\theta$-angle above the threshold \eqref{eq:GRfloor} would be a sign of new CP-breaking physics beyond the Standard Model.

\subsection{Ultraviolet contributions}\label{sec:QCD_UV_contributions}

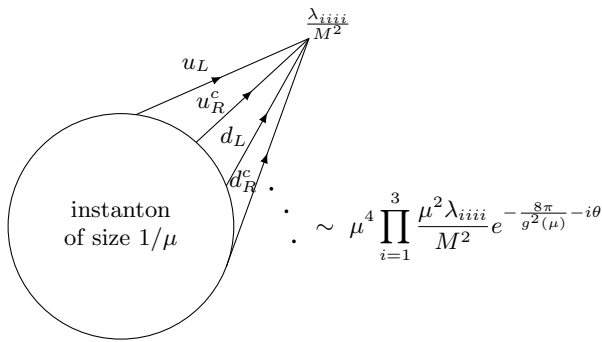
\begin{figure}
	\centering
	\begin{tikzpicture}
	\draw[black](0,0) circle(1.5);
	\begin{scope}[decoration={
		markings,
		mark=at position 0.5 with {\arrow{latex}}}
	]
	\draw[postaction={decorate}] (.2, 1.486) -- (2.5,2.5);
	\draw[postaction={decorate}]  (1, 1.118) -- (2.5,2.5);
	\draw[postaction={decorate}]  (1.4, .538) -- (2.5,2.5);
	\draw[postaction={decorate}]  (1.4, -.538) -- (2.5,2.5);
	\end{scope}
	\node[align=center] at (0,0) { instanton \\ of size $1/\mu$};
	\node[] at (1,2.15) {$u_L$};
	\node[] at (1.2,1.65) {$u_R^c$};
	\node[] at (1.5,1.2) {$d_L$};
	\node[] at (1.63,.6) {$d_R^c$};
	\node[] at (2.75,2.7) {$\frac{\lambda_{iiii}}{M^2}$};
	\node[] at (4.5,0) {$\displaystyle \sim \ \mu^4 \prod_{i=1}^3 \frac{\mu^2 \lambda_{iiii}}{M^2} e^{-\frac{8\pi}{g^2(\mu)}-i\theta}$};
	\node at (2,0.5)[circle,fill,inner sep=0.5pt]{};
	\node at (2.2,.2)[circle,fill,inner sep=0.5pt]{};
	\node at (2.3,-.2)[circle,fill,inner sep=0.5pt]{};
	\end{tikzpicture}
	\caption{Lifting of fermion zero modes by CP-breaking dimension six fermion operator.}
	\label{fig:instanton3}
\end{figure}

Next we estimate the contribution to the axion potential from CP-breaking effects at a high scale $M$.\footnote{We thank Csaba Cs\'{a}ki and Max Ruhrdorfer for useful discussions about this point.} In general, one expects that a gauge
instanton of size $\sim M^{-1}$ will contribute to the axion potential as
\begin{equation}\label{eq:axion_potential_genericUV}
\delta V_{\text{inst.}}^{UV}(M)\sim M^4 e^{i\phi_M}\, e^{-\frac{2\pi}{\alpha_M}-i \theta}+c.c.\, ,
\end{equation}
where $\alpha_M:=\frac{g^2(M)}{4\pi}$, and the phase $\phi_M$ quantifies CP breaking at the scale $M$. Concretely, CP-breaking Standard Model operators of dimension six and higher (as classified in \cite{Buchmuller:1985jz}) can lift all the fermion zero modes.  As an example,
we consider the dimension six operators
\begin{equation}
\mathcal{O}_{6}^{ijkl}:= \epsilon_{ab}D^i Q^{j,a} U^k Q^{l,b}\, ,
\end{equation}
where $a,b=1,2$ index the fundamental representation of $SU(2)_L$.
Insertion of such operators into the effective action as $S_{\text{eff}}\supset \int d^4x\, \lambda_{ijkl}M^{-2}\mathcal{O}_{6}^{ijkl}+c.c.$ with nonvanishing dimensionless Wilson coefficients $\lambda_{ijkl}$ lifts the fermion zero modes of the instanton background: see Figure \ref{fig:instanton3}.
Importantly, in view of assumption (iii) of \S\ref{sec:qcd}, the Wilson coefficients $\lambda_{ijkl}$ will generically have $\mathcal{O}(1)$ \emph{complex} values.

The resulting potential for the QCD axion takes the schematic form
\begin{equation}\label{eq:axion_potential_BSM}
\delta V_{\text{inst.}}^{UV}\sim \int \frac{d\mu}{\mu} \mu^4 \prod_{i=1}^{3}\frac{\mu^2 \lambda_{iiii}}{M^{2}} e^{-\frac{2\pi}{\alpha_\mu}-i\theta}+c.c.\,
\end{equation}
We refer to the term \eqref{eq:axion_potential_BSM} as coming from \emph{high-energy QCD instantons}.
Unless $\prod_{i=1}^3\lambda_{iiii}\in \mathbb{R}$, one indeed obtains a term of the form \eqref{eq:axion_potential_genericUV} with non-trivial CP breaking phase $\phi_M=\text{arg}(\prod_{i=1}^3\lambda_{iiii})$.

The magnitude of \eqref{eq:axion_potential_BSM} is controlled by the properties of QCD at high energies.
Depending on the matter content of QCD, and assuming $\mathcal{O}(1)$ CP breaking in the UV, small gauge instantons may give a contribution to the axion potential that dominates over the IR potential \eqref{eq:axion_potential_QCD}, and thus spoils the PQ solution to the strong CP problem. This problem can be arbitrarily severe if one makes no further assumptions about the QCD spectrum at high energies.

Fortunately, however, this problem of spoliation by high-energy QCD instantons is generally evaded, provided that the charged spectrum is only that of the Standard Model supplemented by superpartners at a SUSY breaking scale $M_{\mathrm{SUSY}}$.
In a supersymmetric theory, a gauge instanton has two universal fermion zero modes associated to the breaking of half the supercharges by the instanton.  These can only be lifted by a SUSY breaking spurion, i.e.~an insertion of the Goldstino mass, which we take to be comparable to the masses of all other superpartners, i.e.~of order $M_{\mathrm{SUSY}}$. The instanton contribution to the scalar potential of \eqref{eq:axion_potential_genericUV} is then replaced by\footnote{We have normalized the instanton potential so that it matches the leading contribution to the F-term potential of the four-dimensional supergravity models considered in this paper. In other models the normalization might be slightly different.}
\begin{equation}\label{eq:axion_potential_SUSYUV}
\delta V_{\text{inst.}}(M)\sim \frac{8\pi}{\alpha_M}M_{\mathrm{SUSY}}M^3 \, e^{-\frac{2\pi}{\alpha_M}-i(\theta-\phi_M)}+c.c.\,
\end{equation}
Matching the running of the QCD gauge coupling (ignoring threshold corrections) one then finds that\footnote{Here, for simplicity we use six `active' quarks down to the Z-boson threshold, ignoring the fact that $m_{top}>m_Z$: this produces only an $\mathcal{O}(1)$ error.}
\begin{equation}\label{eq:highscale_instanton_potential}
\delta V_{\text{inst.}}^{M>M_{\mathrm{SUSY}}}\sim \frac{8\pi}{\alpha_M} \left(\frac{m_Z}{M_{\mathrm{SUSY}}}\right)^3 m_Z^4 e^{-\frac{2\pi}{\alpha_Z}}e^{-i(\theta-\phi_M)}\, ,
\end{equation}
which is independent of the scale $M$, except for the logarithmic scale dependence of $\alpha_M$. Here, $m_Z\approx 91$ GeV is the mass of the $Z$-boson, and $\alpha_Z:=\alpha_{m_Z}\approx 0.119$. Thus, upon ignoring an overall $\mathcal{O}(1)$ factor\footnote{In the extreme case where $M\gtrsim M_{GUT}$ and $M_{SUSY}=1\,$TeV this means accepting a mild correction factor $\alpha_Z/\alpha_M\approx 4$.} $\alpha_Z/\alpha_M$ that depends only logarithmically on the value of $M$ due to the running of $\alpha$ between the mass of the Z-boson and the UV scale $M$, one finds a contribution to the axion potential from high-energy QCD instantons of order $\delta V_{\text{inst.}}\sim 10^{-12}\left(\frac{1\,\text{TeV}}{M_{\mathrm{SUSY}}}\right)^3\Lambda_{QCD}^4$. Such a contribution would shift the minimum of the QCD axion potential to
\begin{equation}\label{eq:theta_vs_Msusy}
\Delta \theta_{\mathrm{QCD}}\sim 10^{-12}\left(\frac{1\,\text{TeV}}{M_{\mathrm{SUSY}}}\right)^3\, .
\end{equation}
Thus, given the fact that $M_{\mathrm{SUSY}}\gtrsim 1$ TeV \cite{Canepa:2019hph}, we find that \eqref{eq:theta_vs_Msusy} is within the experimentally allowed range.

\begin{figure}
\centering
\includegraphics[width=0.8\columnwidth]{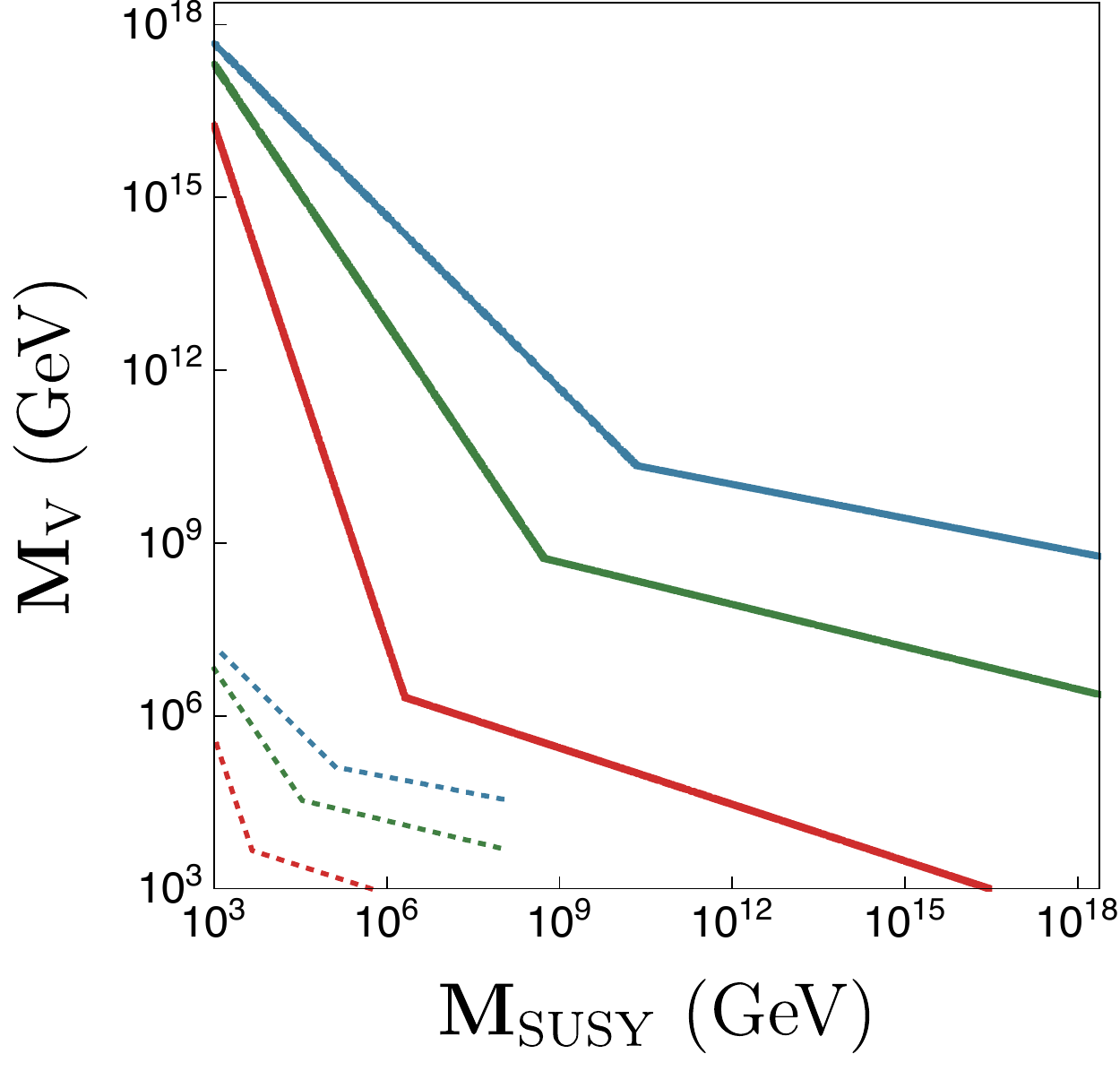}
\caption{Lower bound on the vectorlike masses as a function of the scale of supersymmetry breaking, for UV reference scales $M=M_\mathrm{pl}$ (solid) and $M=10^8$ GeV (dashed). In red is the lower bound for one vectorlike pair, in green for two, and in blue for three. Theories living above these lines have allowed contributions to the $\theta$-angle from high-scale QCD instantons.}
\label{fig:msusy_mvec}
\end{figure}

But importantly, this conclusion is sensitive to our SUSY desert assumptions --- (i) and (ii) of \S\ref{sec:qcd} --- that the QCD spectrum is empty up to high scales except for superpartners. To see this, we now relax our assumptions  (ii)$\rightarrow $(ii$^*$), i.e.~we allow for $n=1, \ 2, \ \text{or} \ 3$ vectorlike pairs (in chiral multiplets) with mass $M_V$ in the window $M>M_V>M_{\mathrm{SUSY}}$.\footnote{Although true vectorlike pairs with small masses are arguably non-generic, string theory realizations of the Standard Model often contain extra Green-Schwarz massive $U(1)$ factors \cite{Cvetic:2011iq,Halverson:2013ska}. Pairs that are vectorlike with respect to the Standard Model can still be chiral with respect to the extra $U(1)$ factors and have naturally small masses generated by instantons. We thank Jim Halverson for pointing this out to us, and for an insightful discussion about this possibility.} This leads to a softer running of the QCD coupling in the UV, and the right-hand side of \eqref{eq:highscale_instanton_potential} gets modified to
\begin{equation}\label{eq:smallinstanton}
	\frac{8\pi}{\alpha_M}\left(\frac{m_Z}{M_{\mathrm{SUSY}}}\right)^3 \left(\frac{M}{M_V}\right)^n m_Z^4 e^{-\frac{2\pi}{\alpha_Z}}e^{-i(\theta-\phi_M)}+c.c.\,
\end{equation}
For opposite ordering of scales $M>M_{\mathrm{SUSY}}>M_V$, and scalar masses of order $M_{\mathrm{SUSY}}$, one obtains \eqref{eq:smallinstanton} with $M_V\rightarrow M_{\mathrm{SUSY}}^{\frac{1}{3}}M_V^{\frac{2}{3}}$. Again, for large SUSY breaking scale and not too light vectorlike pairs, $\Delta \theta_{\mathrm{QCD}}$ is below the experimental bound, but for low SUSY breaking scale, and at least one light vectorlike pair, the PQ mechanism can be spoiled. We show the allowed region in Fig.~\ref{fig:msusy_mvec}.  Under our assumptions, preserving the PQ mechanism implies that a light vectorlike pair observed at collider experiments would imply an unobservably large SUSY breaking scale, and vice versa.

We therefore find, with hardly any model-dependence, that PQ quality is endangered the least when the SUSY desert assumptions hold.
This leads us to a striking conclusion.
If the scale of SUSY breaking is relatively low, $M_{\mathrm{SUSY}}\lesssim 100\,$ TeV, then
the CP-breaking effects of high-energy QCD instantons exceed the
CP breaking \eqref{eq:GRfloor} from weak interactions,
even if the scale $M$ where CP breaking becomes $\mathcal{O}(1)$
is as high as the Planck scale.
Thus, an observable SUSY breaking scale would imply \emph{either} that the QCD $\theta$-angle is greater than the Standard Model expectation computed in \eqref{eq:GRfloor}, \emph{or} that the UV completion preserves CP to a considerable degree.
In the rather optimistic scenario that SUSY is ultimately found close to the TeV scale and future experiments constrain the neutron electric dipole moment below the threshold given by \eqref{eq:theta_vs_Msusy}, one would rule out the entire landscape of solutions considered in this paper.  Ultimately, one would either rule out the hypothesis of $\mathcal{O}(1)$ CP breaking in the deep UV, or reintroduce the strong CP problem, without a viable axion solution.

Having understood the contribution to the axion potential from QCD effects at low and high scales, we now turn to all other contributions to the axion potential that arise in UV-complete models. For this, we will need to invoke concrete UV completions in string theory.

\section{The Peccei-Quinn quality problem in type IIB string theory}
\label{sec:quality}
We will now detail the circumstances in which a model in type IIB string theory possesses a quality problem that obstructs the Peccei-Quinn solution to the strong CP problem. For prior work on this subject, see e.g. \cite{Svrcek:2006yi,Conlon:2006tq,Cicoli:2012sz,Bachlechner:2019vcb,Broeckel:2021dpz,Cicoli:2021gss}.

\subsection{Type IIB conditions for a PQ quality problem} \label{IIBPQ}

To begin, we will briefly describe features of low-energy supergravity theories that arise via compactification of type IIB string theory on Calabi-Yau orientifolds.  We will focus on expressing the instanton generated potential \eqref{eq:UVpotential} in terms of geometric data.

In type IIB on a Calabi-Yau orientifold\footnote{We do not perform an explicit orientifold, but we note that projecting instantons out by orientifolding would only strengthen our results on PQ quality.} $X$ (of O3/O7 type) with fluxes, or an F-theory generalization thereof, the effective theory  at low energies is well-described by $\mathcal{N}=1$ supergravity with Standard-Model-like
gauge sectors from open string degrees of freedom on seven-branes and their intersection curves (vector multiplets and charged chiral multiplets, respectively), along with $N$ gauge singlet chiral multiplets $T_A$, $A=1,\ldots,N$, where $N = \mathrm{dim}\, H^2(X,\mathbb{R})$. The real parts $\tau_A:=\text{Re}(T_A)$ are geometric moduli of the threefold $X$ and can be thought of as the volumes of a homology basis of four-cycles $D_A\in H_4(X,\mathbb{Z})$, and the imaginary parts $\xi_A:=\text{Im}(T_A)$ are the axion zero modes of the four-form potential $C_4$.
The (classical) holomorphic Yang-Mills coupling associated with a stack of seven-branes wrapped on a divisor in a class $q^A [D_A]$ is equal to $q^A T_A$, and the linear combination $q^A \xi_A$ takes on the role of a dynamical $\theta$-angle.
The gauge symmetry $C_4\rightarrow C_4+d\Lambda_3$ in ten dimensions implies that the $\xi_A$ receive no scalar potential to any perturbative order in the large volume expansion. Rather, their potential is entirely generated by instantons: specifically, by
Euclidean D3-branes wrapped on 4-cycles in the compactification manifold.

The action of a Euclidean D3-brane wrapping an \textit{effective} divisor $\Sigma_{\vec{q}}$ (i.e. a holomorphic four-cycle) in a homology class $[\Sigma_{\vec{q}}] = q^A [D_A]$ is
\begin{align}
S_E^{\vec{q}}=2 \pi \operatorname{Vol}\left(\Sigma_{\vec{q}}\right)=2\pi q^A \tau_A\, ,
\end{align}
and the instanton expansion of the superpotential is, setting $M_\mathrm{pl}=1$,
\begin{align}\label{wis}
W=W_{0}+\sum_{\vec{q}} \mathcal{A}_{\vec{q}} \,\exp \left(-2 \pi q^{A} T_{A}\right)\, .
\end{align}
Here, $W_0$ is a (generally complex) constant term generated by fluxes, and the sum runs over all effective divisors. Euclidean D3-branes wrapping cycles that are \emph{not} effective break all supercharges, and therefore may at most contribute  to the K\"ahler potential
\begin{align}\label{eq:Kahlerpotential}
K=K_{\text{tree}}(\tau_A)+K_{\mathrm{pert.}}(\tau_A)+K_{\mathrm{np}}(T_A,\overline{T}_{\bar{A}})\, .
\end{align}
Here, the tree-level K\"ahler potential is $K_{\text{tree}}(\tau_A)=-2\log \mathcal{V}$ where $\mathcal{V}$ is the geometric volume of $X$ (as a function of the $\tau_A$), and the perturbative and non-perturbative corrections $K_{\mathrm{pert.}}(\tau_A)$ and $K_{\mathrm{np}}(T_A,\overline{T}_{\bar{A}})$ are not known in general, but are negligible when the curvature of the internal space is small.

For simplicity, let us assume that the $\tau_A$ are stabilized, perturbatively, at a point in field space where $\mathcal{A}_{\vec{q}}\,e^{-2\pi q^AT_A}\ll|W_0|\ll 1$. Then the F-term potential can be written as
\begin{equation}
V_F=V_{\text{pert.}}(\tau_A)+V_{\text{axion}}(\tau_A,\xi_A)\, ,
\end{equation}
with
\begin{align}
V_{\text{axion}}(&\tau_A,\xi_A)=2m_{\frac{3}{2}}e^{\frac{K}{2}}\text{Re}\left(\bar{\partial}^AK_{\text{pert.}}\partial_A W_{\text{np}}+\epsilon \, W_{\text{np}} \right)\nonumber\\
&+\mathcal{O}\Bigl(W_0^2e^{-2\pi q^A\tau_A},e^{-2\pi(q^A+{q'}^A)\tau_A}\Bigr)\, ,
\end{align}
where $m_{\frac{3}{2}}:=e^{\frac{K}{2}}|W|$ is the gravitino mass, $\epsilon:= K^{A\bar{B}}K_A K_{\bar{B}}-3$,
and without loss of generality we have taken $W_0\in \mathbb{R}$. In Calabi-Yau compatifications the tree-level K\"ahler potential $K_{\text{tree}}$ satisfies the relations $\epsilon=0$ and $\bar{\partial}^AK_{\text{tree}}=-2\tau^A$, so it is justified to approximate the axion potential \eqref{eq:UVpotential} using
\begin{equation}\label{eq:nonperturbative_scales}
\Lambda_{\vec{q}}^4 \approx 8\pi\,  m_{\frac{3}{2}}\frac{\vec{q}\cdot \vec{\tau}}{\mathcal{V}}\, |\mathcal{A}_{\vec{q}}|\, e^{-2\pi \vec{q}\cdot \vec{\tau}}\, ,
\end{equation}
i.e.~one considers the axion potential generated by irreducible BPS instantons. Note that multi-instanton terms are strictly subleading to terms of the form \eqref{eq:nonperturbative_scales}---therefore, a quality problem introduced by instantons of the form \eqref{eq:nonperturbative_scales} cannot be remedied by the inclusion of multi-instanton terms, and conversely if no quality problem is present at the level of \eqref{eq:nonperturbative_scales}, one cannot be introduced by these subleading effects. Crucially, for any point in field space where the above approximations hold we can compute the leading instanton corrections to the scalar potential up to the standard $\mathcal{O}(1)$ uncertainty in the holomorphic one-loop Pfaffians $\mathcal{A}_{\vec{q}}$. Even if some or all of the instanton terms instead dominate over a small constant term $W_0$, but remain small themselves, on general grounds one may still use the estimate $\Lambda_{\vec{q}}^4\sim m_{\frac{3}{2}}/\mathcal{V} \,\times e^{-2\pi \vec{q}\cdot \vec{\tau}}$ except that $m_{\frac{3}{2}}$ itself is dominated by instanton contributions. Thus, as long as the scales $\Lambda_{\vec{q}}$ in \eqref{eq:nonperturbative_scales} are expressed in terms of the gravitino mass instead of the value of the classical flux superpotential $W_0$, we may estimate the axion potential using \eqref{eq:nonperturbative_scales}.

On the other hand, if the overall superpotential is large, non-BPS instantons can dominate over BPS instantons, and no general methods now exist for computing the axion potential
(see however \cite{Demirtas:2019lfi,Long:2021lon} and Appendix \ref{app:subleading} for comments on this regime). For this reason, we will by necessity take the SUSY breaking scale to be small.

One might suspect that the assumption of a low SUSY breaking scale, made here for reasons of computational control, severely limits what can be said about the prevalence of a PQ quality problem in our framework. However, as we have seen in the previous section, under some mild assumptions about the matter spectrum of QCD, the PQ quality problem tends to get more severe the \textit{lower} the SUSY breaking scale. Thus, we see little reason to expect that the PQ quality problem is present when the SUSY breaking scale is high. In other words, it appears as a fortunate coincidence that one can draw fairly general conclusions about the strong CP problem by studying instanton corrections to the \emph{superpotential} (which is comparatively accessible by direct computation), because the problem becomes milder as one leaves the regime where superpotential terms are dominant.

\subsection{Generating models}\label{sec:generating_models}

In order to assess PQ quality in a \textit{fully explicit} landscape of models of the above type one would have to generate a rather impressive set of data:
\begin{enumerate}
	\item A large set of Calabi-Yau threefolds.
	\item One or more consistent O3/O7 orientifold involutions for each threefold.
	\item For each geometric model, choices of seven-branes and fluxes that generate a realistic Standard Model, with somewhat small scale of SUSY breaking, moduli stabilization, the correct dark matter abundance, etc.
\end{enumerate}
For reasons of time we will not pursue achieving this here. Instead, we will content ourselves with a large set of Calabi-Yau threefolds from the Kreuzer-Skarke list \cite{Kreuzer:2000xy}; a set of effective divisors that could in principle host QCD; and a point in the geometric moduli space of each Calabi-Yau where the instanton expansion of the superpotential converges and the QCD gauge coupling comes out right.
We do not know which, if any, of these toy models can actually be completed to fully realistic models of particle physics and cosmology.
Even so, it
would require a quite remarkable coincidence for the divisors hosting QCD, and the points in moduli space where moduli stabilization \textit{actually} occurs to be so special as to invalidate conclusions drawn from the statistics of a large random sample.

A few words are in order about the construction of our ensemble.
Our aim is to generate a random sample of Calabi-Yau manifolds, and determine an upper bound on the QCD $\theta$-angle in these geometries. We will realize these manifolds as hypersurfaces in toric varieties, following \cite{Batyrev}. The starting point is a four-dimensional reflexive polytope $\Delta^\circ$ from the Kreuzer-Skarke list and its set of fine regular star triangulations
$\{\mathcal{T}\}$, each defining a generically singular toric fourfold $V_{\mathcal{T}}$ whose generic anti-canonical hypersurface defines a smooth Calabi-Yau threefold $X_{\mathcal{T}}$.\footnote{In this work, we will restrict our ensemble to favorable hypersurfaces for ease of computation. We are not aware of any properties of non-favorable hypersurfaces that would modify our results.}

Let $\mathcal{K}_{\mathcal{T}}$ be the sub-cone of the K\"ahler cone of $X_\mathcal{T}$ that is inherited via restriction from the K\"ahler cone of $V_{\mathcal{T}}$, and denote by $\mathcal{K}:=\cup_{\mathcal{T}}\mathcal{K}_{\mathcal{T}}$ the \textit{inherited extended K\"ahler cone}. Intuitively, $\mathcal{K}$ can be thought of as the part of the Calabi-Yau threefold's K\"ahler moduli space that is describable via embedding into the set of birationally equivalent toric fourfolds.

We obtain a random sample of triangulations using Algorithm $\# 3$ of \cite{Demirtas:2020dbm}. An implementation of this algorithm is included in \texttt{CYTools} \cite{CYTools}, via the function \texttt{random\_triangulations\_fair}: we used the parameters \texttt{initial\_walk\_steps=500}, \texttt{n\_walk=50} and \texttt{n\_flip=50}.

This method picks out a random Calabi-Yau compactification. Equipped with an explicit compactification manifold, one might expect that all our conclusions will heavily depend on the precise choice of point in its K\"ahler moduli space. For each direction in the K\"ahler cone, the overall dilation is fixed by demanding that QCD has the right gauge coupling (to be described in detail momentarily), so our results could, a priori, be sensitive to the choice of ray in $\mathcal{K}_{\mathcal{T}}$. However, at large $h^{1,1}$, the inherited extended K\"ahler cone contains exponentially many phases $\mathcal{K}_{\mathcal{T}}$ connected to each other via bi-stellar flips in the triangulation.  As a consequence, within each individual phase $\mathcal{K}_{\mathcal{T}}$, divisor volumes are expected to depend only very weakly on the choice of direction in the K\"ahler cone.

Thus, for simplicity, in each phase we select as reference point a homogeneous dilation of the \textit{tip of the stretched K\"ahler cone}, $t_{\mathrm{tip}}$, which has been used in various recent studies of the string landscape \cite{Demirtas:2018akl,Halverson:2019cmy,Demirtas:2020dbm,Mehta:2020kwu,Mehta:2021pwf}, and is defined as the point closest to the origin where all effective curves have volumes $\geq 1$. As explained above, we do not expect this choice to actually matter, but in order to test our expectation quantitatively we have run auxiliary scans using other randomly chosen rays in $\mathcal{K}_{\mathcal{T}}$, obtained via a slight variation of Algorithm $\# 2$ of \cite{Demirtas:2020dbm}, with \texttt{n\_walk=10}. We indeed find that our results remain the same to good approximation.

Having chosen a one-dimensional ray for each geometric model, we now impose two important constraints:
\begin{itemize}
\item The gauge coupling of QCD at this point must match the gauge coupling of QCD in the IR.
\item The instanton expansion must be under control.
\end{itemize}
In order to address the first point, we note that along \textit{any} ray in $\mathcal{K}_{\mathcal{T}}$ and any choice of divisor hosting QCD there exists a point at which it would have the correct volume to reproduce the QCD gauge coupling at low energies.
Thus, we need to (a) choose a divisor hosting QCD and (b) homogeneously scale $t_{\mathrm{tip}}$ in order to ensure that the high-scale QCD coupling indeed runs to the observed QCD coupling at low energies. For step (a) the toric fourfold provides natural candidates: the intersections of the $h^{1,1}(V)+4$ toric divisors of $V$ with the Calabi-Yau hypersurface. We will generate a model for a sample of up to five of these. We emphasize that step (b) is sensitive to the running of the QCD coupling in the UV, which, under assumptions (i) and (ii$^*$), is set by the SUSY breaking scale, the number $n$ of vectorlike pairs charged under QCD, and their mass $M_V$.

The correct homogeneous scaling of the K\"ahler parameters can be determined from the running of the QCD coupling from the UV matching scale $M$ to a low energy scale, and must be chosen such that the following equation holds,
\begin{align}\label{eq:matching_condition}
	e^{-2\pi \text{Vol}(D_Q)}=& \left( \frac{m_Z}{M_{\mathrm{SUSY}}}\right)^3   \left(\frac{M}{\accentset{\sim}{M}_V}\right)^n   \nonumber \\ & \times \frac{m_Z^4}{M^3 M_{\mathrm{SUSY}}}e^{-\frac{8\pi^2}{g^2(m_{Z})}}\, ,
\end{align}
where $D_Q$ is the divisor hosting QCD, and $\accentset{\sim}{M}_V:=M_V$ for $M_V>M_{\mathrm{SUSY}}$ and $\accentset{\sim}{M}_V:=M_V^{\frac{2}{3}}M_{\mathrm{SUSY}}^{\frac{1}{3}}$ otherwise. For a given ray in the K\"ahler cone, this is a transcendental equation in one real parameter, which is straightforward to solve numerically. In what follows we will primarily focus on the case $n=0$ and comment on the generalization to $n>0$ at the end of \S\ref{sec:results}. Importantly, as PQ quality is endangered most for small SUSY breaking scale, we will set the masses of superpartners to be of order $1$ TeV $\approx 10^{-15} M_\mathrm{pl}$, roughly as light as allowed with current exclusion limits from the LHC.

After performing the homogeneous rescaling to impose \eqref{eq:matching_condition}, we compute the volumes of all prime toric
divisors (cf.~Appendix \ref{app:subleading}) in the Calabi-Yau. In order to impose our second constraint, i.e.~convergence of the instanton expansion, we will not attempt to assess PQ quality in models with small divisors: such theories are not under computational control, and statements about the $\theta$-angle based on the first few terms in a non-convergent series expansion should not be trusted. Thus, once we have dilated the overall volume to match the QCD gauge coupling at low energies, we will only compute the $\theta$-angle \textit{if} all divisor volumes are greater than one:
\begin{align}
	\mathrm{Vol}(D_I) \geq 1  \ \ \ \ \forall \ I.
	\label{eq:divstretch}
\end{align}
It would clearly be of great interest to repeat the analysis in the regime of small divisor volumes, but such a computation is beyond  the scope of this work. One could  speculate that such an investigation
would reveal that Peccei-Quinn quality is positively correlated with control of the $\alpha'$ expansion, i.e.~that the strong CP problem is solved in (most of) the geometric regime, and \emph{not} solved in large portions of the \emph{non}-geometric regime.  Such a finding
would be a rare example of more desirable phenomenology occurring in a regime that is convenient for theorists.  Indeed, this would be an example of the proverbial keys actually being under the lamppost.

Finally, we need to define what precisely we mean by the UV scale $M$. First, we note that in general it does not correspond to a physical mass threshold but is rather a generally unphysical matching scale chosen such that extrapolation of the four-dimensional effective theory to that scale gives $\frac{4\pi}{g^2(M)}=\text{Vol}(D_Q)$. It is straightforward to estimate this scale by comparing to, for instance, the case of gaugino condensation in pure super-Yang-Mills obtained from $\mathfrak{c}$ seven-branes on a rigid divisor $D$. In the supergravity theory one expects a superpotential $W_{np}\supset \mathfrak{c} \mathcal{A}_{D}e^{-2\pi T_D}$ where $T_D$ is the complexified volume of the cycle, and $\mathcal{A}_D$ is a one-loop Pfaffian that depends only on complex structure moduli and the dilaton, and thus generically takes $\mathcal{O}(1)$ values. The well-known gaugino condensation superpotential in four-dimensional super Yang-Mills is $\mathfrak{c} \Lambda^3$ \cite{Veneziano:1982ah,Affleck:1983mk} where $\Lambda^3=M^3 e^{-2\pi T/\mathfrak{c}}$ and this gets identified with $e^{\frac{K}{2}}M_\mathrm{pl}^3\,\mathfrak{c} \mathcal{A}_{D}e^{-2\pi T_D}$ where $K=-2\log(\mathcal{V})-\log(1/g_s)$ is the K\"ahler potential. We thus find
\begin{equation}
	M\simeq e^{\frac{K}{6}}M_\mathrm{pl}=\frac{g_s^{\frac{1}{6}}}{\mathcal{V}^{\frac{1}{3}}}M_\mathrm{pl}=M_s^{\frac{2}{3}}M_\mathrm{pl}^{\frac{1}{3}}\,,
\end{equation}
where $M_s$ is the string scale. Note that $M_{KK}<M_s<M<M_\mathrm{pl}$, so the matching scale $M$ is actually larger than the na\"ive cutoff of the four-dimensional effective theory, namely the Kaluza-Klein scale. For this work, we will assume only moderately small string coupling, so we may choose $M\simeq M_\mathrm{pl}/\mathcal{V}^{\frac{1}{3}}$.

\section{Results} \label{sec:results}

We generated an
ensemble of $32{,}040$ manifolds with Hodge numbers ranging from $h^{1,1} =2$ to $h^{1,1} = 491$, with up to $10$ polytopes per $h^{1,1}$ and up to $10$ triangulations per polytope, along with the values of the K\"ahler moduli at the tip of the stretched K\"ahler cone in each geometry. For each such point, we considered hosting QCD on up to $5$ prime toric divisors.\footnote{For some values of $h^{1,1}$ there are fewer than $10$ polytopes, and for some polytopes there are fewer than $10$ triangulations.  The number of models per Calabi-Yau can be less than five if there are not enough divisors whose volumes can generate the correct QCD gauge coupling via an overall dilation without shrinking any other divisors to have volumes less than one.} As stated in \S\ref{sec:generating_models}, we chose the scale of supersymmetry breaking to be $1$ TeV in order to put the Peccei-Quinn mechanism in the most danger.
In total, our ensemble consists of 136{,}659 distinct models.

\begin{figure*}
\centering
\includegraphics[width=1.8\columnwidth]{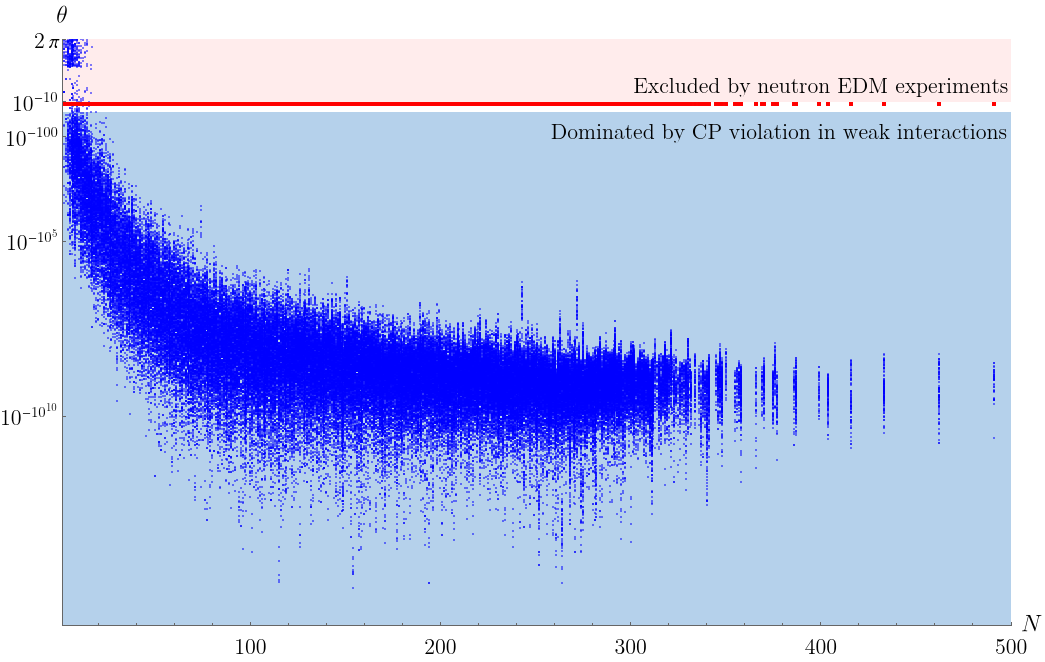}
\caption{Contributions to the $\theta$-angle as functions of the number of axions, $N$.  For each model --- a specific Calabi-Yau with a choice of a divisor hosting QCD --- there is a red point showing $\Delta \theta_{\mathrm{QCD}}$ and a blue point showing $\Delta \theta_{\mathrm{stringy}}$, cf.~\eqref{eq:deltathetattot}. $136{,}659$ models are shown. The pink region indicates the region $\theta > 10^{-10}$, which is excluded by bounds on the neutron electric dipole moment. The blue region shows the parameter space $\theta <10^{-19}$, which is below the floor set by CP-breaking effects in the weak sector of the Standard Model.}
\label{scan}
\end{figure*}

In Fig.~\ref{scan}, $\Delta \theta_{\mathrm{stringy}}$ and $\Delta \theta_{\mathrm{QCD}}$ are plotted as functions of the number of axions $N\equiv h^{1,1}$, in blue and red, respectively. Each point represents one model.
Recall that the total upper bound on the $\theta$-angle is
\begin{align}
\Delta \theta = \Delta \theta_{\mathrm{QCD}}+ \Delta \theta_{\mathrm{stringy}}\,,
\end{align}
as in \eqref{eq:deltathetattot}.

We draw the reader's attention to two regions on this plot.  The region shaded red is excluded by experimental bounds on the neutron electric dipole moment. The fact that $\theta$ is found not to be in this region is the strong CP problem.  In the region shaded blue, the $\theta$-angle is smaller than the floor \eqref{eq:GRfloor} set by CP-breaking in the weak sector \cite{Georgi:1986kr}.  In the narrow band between the red and blue regions, one could hope to measure a $\theta$-angle that is not currently excluded, but is governed by physics beyond the Standard Model.

In a small fraction of the models with $N \lesssim 20$, $\Delta \theta_{\mathrm{stringy}}$ is large enough to spoil the PQ mechanism: specifically,
of the $6{,}716$ models in the ensemble with $N\leq 17$, $414$ (8\%) have $\Delta \theta_{\mathrm{stringy}}>10^{-10}$.
However, $\Delta \theta_{\mathrm{stringy}}$ quickly becomes utterly negligible as $N$ increases: we find
\begin{align}\label{eq:thetastringy}
\Delta \theta_{\mathrm{stringy}} \propto e^{-c N^{4}}\, ,
\end{align}
with $c \approx 1.8$.

We therefore conclude
\begin{equation}\label{theboxed}
\boxed{\phantom{\Biggl(\Biggr)}\Delta \theta_{\mathrm{stringy}} \ll \Delta \theta_{\mathrm{QCD}} < 10^{-10}\, .~~~~}
\end{equation}
The first inequality in \eqref{theboxed} holds for $N \gg 1$, but is independent of the number $n$ of vectorlike pairs. The second inequality is independent of $N$ but is modified if $n>0$ (see \S\ref{sec:QCD_UV_contributions}).

Fig.~\ref{scan} shows that for $N >17$, $\Delta \theta_{\mathrm{QCD}}$ is the dominant contribution to the $\theta$-angle. \footnote{For small values of $N$, it is sometimes the case that $\Lambda_{\vec{q}_c}^4 \gg \Lambda_{\mathrm{QCD}}$. In these cases, we select a random value for $\Delta \theta_{\mathrm{stringy}}$ between $0$ and $2\pi$.}
Notably, at $M_{\mathrm{SUSY}}=1$ TeV, $\Delta \theta_{\mathrm{QCD}}$ is larger than the floor \eqref{eq:GRfloor} from CP-breaking by the weak interactions.
Even so, as explained in \S\ref{sec:QCD_UV_contributions}, \emph{in the absence of vectorlike pairs} the contribution $\Delta \theta_{\mathrm{QCD}}$ is always below the experimental bound on $\theta$, and the only effect that can shift the minimum of the QCD axion potential far enough to spoil the PQ mechanism is $\Delta \theta_{\mathrm{stringy}}$.

We conclude that theories in this landscape with $N \gtrsim 20$ axions and no vectorlike pairs
are  exponentially likely to realize the Peccei-Quinn solution of the strong CP problem. This is one of our main results.

We end with two remarks.
First, we comment on the quality problem in the case that the non-perturbative effects that contribute to \eqref{wis} are not Euclidean D3-branes, but rather strong gauge dynamics on D7-branes stacked on various cycles.
Gaugino condensation on a four-cycle with volume $T$ wrapped by $\mathfrak{c}$ D7-branes would produce a contribution $\propto e^{-2\pi T/\mathfrak{c}}$ to \eqref{wis}, thereby enhancing the scale of $V_{\mathrm{hidden}}$.  If $\mathfrak{c}$ could be arbitrarily large, then this enhancement could be great enough to spoil the PQ mechanism. However, in a typical model, one might expect to find $\mathfrak{c} \sim 6$ --- see e.g.~\cite{Denef:2005mm,Demirtas:2021nlu}.  For any comparably small
value of $\mathfrak{c}$, the conclusions we obtained about the Peccei-Quinn mechanism remain
unchanged: the values of $\Delta \theta_{\mathrm{stringy}}$ that we found are so small (see Fig.~\ref{scan}) that the qualitative conclusions are robust to $\mathcal{O}(1)$ changes in divisor volumes.  Even so, the precise value of $N$ for which good quality is guaranteed may shift. Furthermore, if $\mathfrak{c}$ were large, it would appear reasonable to require that all divisor volumes are $\gtrsim \mathfrak{c}$ instead of imposing \eqref{eq:divstretch}, effectively removing the parameter $\mathfrak{c}$ from the problem.

Second, if there are vectorlike pairs charged under QCD, some portions of the space parameterized by the
SUSY breaking scale and the masses of vectorlike pairs become excluded,
as explained in \S\ref{sec:QCD_UV_contributions}. Even so, \textit{provided} these parameters take values in the allowed regime shown in Figure \ref{fig:msusy_mvec}, our results obtained under the assumption of no vectorlike pairs immediately carry over: for each model a change in the above parameters simply translates into a dilation of the overall volume via the matching condition \eqref{eq:matching_condition}, and the relative cycle sizes are invariant. Therefore, if stringy instantons are subleading to the high scale QCD instantons in a model with no vectorlike pairs, they will remain so in a corresponding model with $n>0$ pairs.

\subsection{Cosmological and astrophysical bounds}

Up to this point we have analyzed the strong CP problem in an ensemble of axion theories, without imposing experimental constraints other than the upper bound on the neutron electric dipole moment and the lower bound on the scale of supersymmetry breaking.
We now consider constraints on this ensemble from the measured dark matter relic density, and from stellar emission of axions and axion-like particles (ALPs).
Such limits are not the main purpose of this work, and so our treatment will be schematic, and follow well-known lines. A comprehensive analysis would be a useful target for future work.

We begin by discussing the physics considerations that lead to a bound on the axion dark matter relic density.
We assume that the initial values of the dimensionless axion fields are of order unity, and we consider each species that is stable on cosmological timescales, i.e.~with (cf.~e.g.~\cite{Marsh:2015xka})
\begin{align}
\frac{\tau_{\text{ALP}}}{\tau_{\text{universe}}} = 10^{-15}\left(\frac{\text{GeV}}{m_a}\right)^3 \left(\frac{f_a}{10^{12} \ \text{GeV}}\right)^2 \gtrsim 1\,,
\label{lifetime}
\end{align}
where $\tau_{\text{universe}}$ is the age of the universe, $m_a$ is the ALP mass, and $f_a$ is the ALP decay constant.
The fractional energy density of one\footnote{We do not incorporate mixings of axion species, though it would be interesting to do so. For studies of effects of axion mixing on dark matter abundance and on axion-photon conversion, see \cite{Cyncynates:2021yjw} and \cite{Chadha-Day:2021uyt}, respectively.} such ALP is \cite{Marsh:2010wq}:
\begin{align}
\Omega_a \approx \frac{1}{6} (9 \Omega_r)^{3/4} \left(\frac{m_a}{H_0}\right)^{1/2}  \left(\frac{\phi_i}{M_\mathrm{pl}}\right)^2\,,
\end{align}
where $\Omega_r$ is the fractional energy density due to radiation, and $\phi_i$ is the initial axion field displacement. The observed relic energy density of dark matter $\Omega_{DM} h^2 = 0.12$ then leads to a bound on the ALP mass:
\begin{align}
m_a \lesssim \frac{10^{37} \ \text{GeV}^5}{f_a^4}\,.
\label{relic}
\end{align}
We will consider a model to be cosmologically viable if all the axions in the theory that satisfy \eqref{lifetime} also   satisfy \eqref{relic}.

For the specific case of the QCD axion, one can use $m_a \sim \frac{m_\pi f_\pi}{f_a}$ in \eqref{relic} to find the standard bound
\begin{align}
f_a \lesssim 10^{12} \ \text{GeV}
\end{align}
on the QCD axion decay constant, for an initial misalignment angle of order unity.

Observational constraints on the axion-photon coupling, $g_{a\gamma \gamma}$, provide lower bounds on the QCD axion decay constant.  One of the most stringent bounds comes from the CERN Axion Solar Telescope (CAST) \cite{CAST:2017uph}, which finds
\begin{align}
g_{a\gamma \gamma} \lesssim 0.66 \times 10^{-10} \ \text{GeV}^{-1}
\end{align}
for axions with masses less than $0.02$ eV.\footnote{Other axion searches are sensitive to lower values of the axion-photon coupling in other regions of parameter space \cite{Wouters:2013hua, Berg:2016ese, Marsh:2017yvc, Conlon:2017qcw, Chen:2017mjf, Reynolds:2019uqt}, but we leave a detailed analysis of these bounds to future work.}

We estimate the coupling $g_{a\gamma \gamma}$ by
taking the axion that couples to $F_{\mu\nu} F^{\mu \nu}$ to be a general linear combination of all mass and kinetic eigenstate axions in the theory, with $\mathcal{O}(1)$ coefficients. Then we define
\begin{align}
g_{a\gamma\gamma, \ \mathrm{eff}} = \frac{\alpha_{\mathrm{EM}}}{2 \pi } \sqrt{\sum_{i|m_i<0.02 \ \text{eV}} \frac{1}{f_i^2}}
\end{align}
and compare this with observational bounds.

\begin{samepage}

\begin{figure}
\centering
\includegraphics[width=\columnwidth]{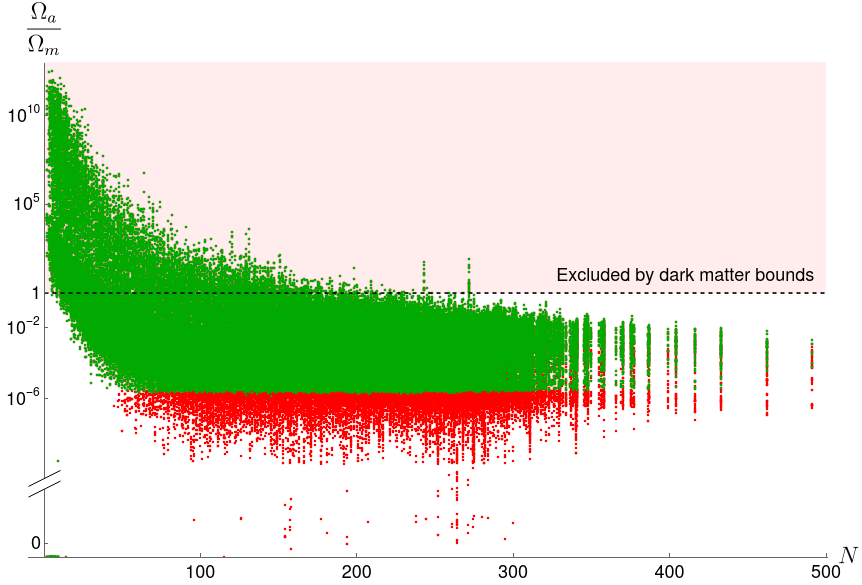}
\caption{The fractional abundance of axion dark matter as a function of $N$. The dotted line at $1$ indicates the place where axion dark matter constitutes the entirety of the observed dark matter abundance. Values higher than this are excluded.
Green points are allowed by the constraints from CAST, while red points ($17\%$) are not.}
\label{dmdensity}

\bigskip

\includegraphics[width=\columnwidth]{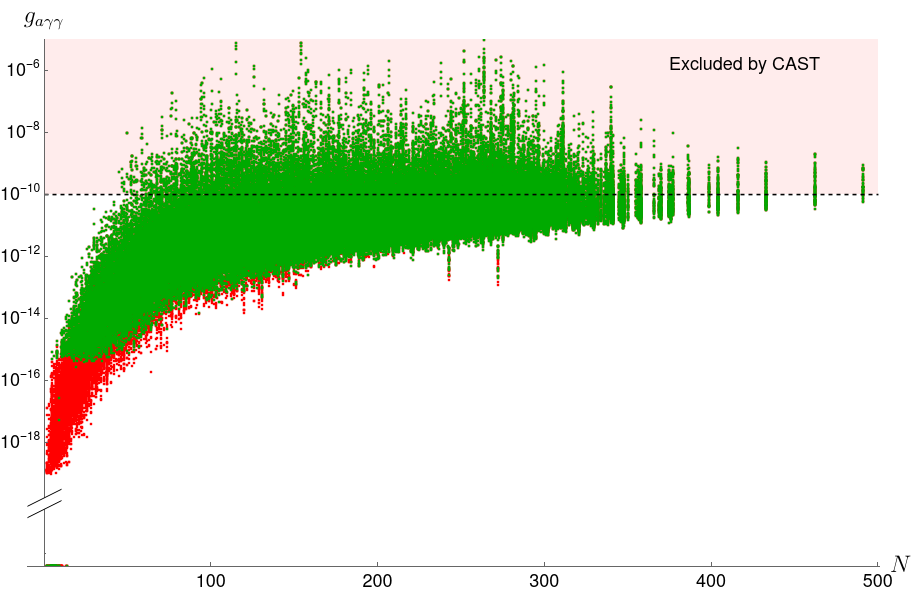}
\caption{The effective axion-photon coupling as a function of $N$. Points above the line are excluded by CAST \cite{CAST:2017uph}.
Green points are allowed by dark matter relic abundance constraints, while red points ($9.4\%$) are not.}
\label{astrobound}

\bigskip

\includegraphics[width=\columnwidth]{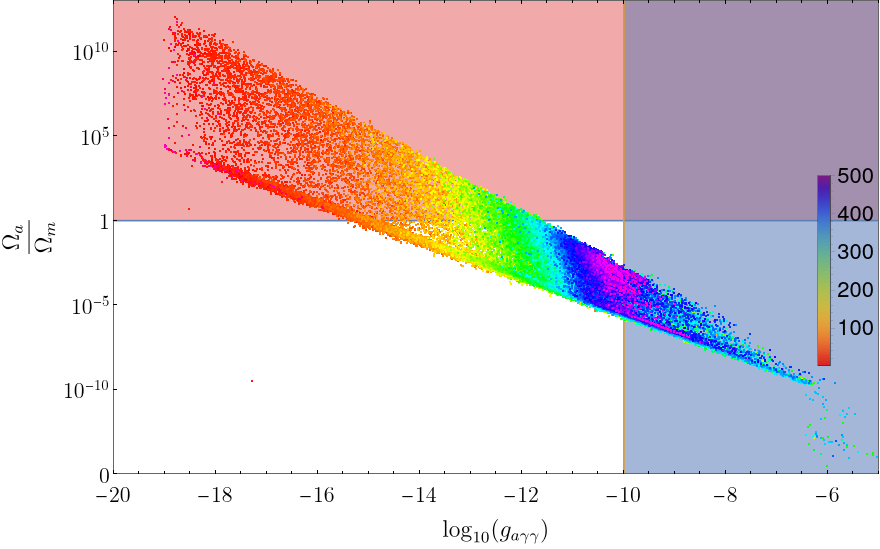}
\caption{Dark matter relic density versus effective $g_{a\gamma\gamma}$, color-coded by $N$. Red points correspond to low $N$, while purple points correspond to large $N$.}
\label{dm_v_astro}
\end{figure}

\end{samepage}

Cosmological and astrophysical constraints on our ensemble are shown in Figs.~\ref{dmdensity}-\ref{dm_v_astro}. Fig.~\ref{dmdensity} shows the fractional abundance of axion dark matter as a function of $N$. Fig.~\ref{astrobound} shows the effective $g_{a\gamma\gamma}$ as a function of $N$. Finally, Fig.~\ref{dm_v_astro} displays these data as dark matter relic density versus effective $g_{a\gamma\gamma}$, color-coded by $N$.\footnote{Roughly $0.1\%$ of points in Figs.~\ref{dmdensity} and \ref{dm_v_astro} have been been omitted for the sake of visual clarity: in these models, the overall scale is so low that the QCD axion itself does not satisfy \eqref{lifetime}, and thus does not contribute to the dark matter density. These models have negligible dark matter abundances, and are not phenomenologically viable---in any case, they violate \eqref{astrobound}.}

Several remarks about these results are in order. First, one can see that an order-one fraction of all the geometries studied are allowed by observational bounds. The dark matter bounds become less stringent as $N$ increases: this is because the Kaluza-Klein cutoff drops with $N$, so decay constants decrease as well. Conversely, astrophysical bounds become more stringent as $N$ increases.  Even so, the models in our landscape are generally not ruled out on astrophysical grounds: the bounds from CAST only come into contact with our data for the largest values of $N$. Second, sharp edges are visible in Fig.~\ref{dm_v_astro}: the lower edge marks the contribution to the dark matter abundance from the QCD axion itself, and therefore constitutes a lower bound on the total dark matter abundance for any given $g_{a\gamma \gamma}$. The upper edge of the distribution is set by the largest-mass axions in each theory, which in turn is dictated by the smallest volume allowed for divisors.
This minimum volume was set to one in \eqref{eq:divstretch}, but if the minimum volume were decreased by some amount, the top edge would marginally increase.

It is interesting to note the interplay of the dark matter relic densities with the $\theta$-angle. We saw above that at as the number of axions $N$ in the effective theory increases, the stringy contribution to the $\Delta \theta$ decreases.  By the same token, as $N$ increases, the axion dark matter abundance decreases. Thus, we find a correlation between the viability of the Peccei-Quinn mechanism and the observed dark matter abundance. This  hints at an anthropic solution of the strong CP problem (this possibility was also discussed in \cite{Dine:2018glh}).

\section{Conclusions}

\label{sec:conclusions}

String theory and experiment are separated by a vast hierarchy of energy scales.
The decoupling of ultraviolet physics from infrared phenomena makes science possible, but at the same time it presents an obstacle to making useful predictions from string theory.
Fortunately, some mysteries of low-energy physics are sensitive to the details of the high-energy theory. The strong CP problem is a leading example: the most promising solution,
the Peccei-Quinn mechanism, can be spoiled by corrections from Planck-suppressed operators.
Explaining the smallness of the neutron electric dipole moment following Peccei and Quinn thus requires understanding quantum gravity effects.

In weakly-coupled string compactifications, one can directly calculate the QCD axion potential, including the effects of the ultraviolet completion of gravity: among these the most important are D-brane instantons.
We computed the leading D-brane instantons in type IIB compactifications on orientifolds of Calabi-Yau threefold hypersurfaces in toric varieties, in the geometric regime.  We found that in this setting, D-brane instanton contributions to the neutron electric dipole moment fall exponentially with the fourth power of the number of axions: see \eqref{eq:thetastringy}. This realizes Polchinski's proposal \cite{Polchinski:2006gy} that the strong CP problem might admit a natural solution in
topologically complex compactifications of string theory.

Assessing the QCD axion potential in these models also required understanding QCD instantons at high energies, as a question in field theory.  We noted a surprising fact: although the high-energy QCD instanton contributions are small enough not to spoil the PQ mechanism in the absence of vectorlike pairs, for a low enough scale of supersymmetry breaking they make contributions to the $\theta$-angle of a size that may be detectable with future experiments.

In the same ensemble, we carried out a preliminary analysis of the axion dark matter relic abundances, as well as effective axion-photon couplings, and compared these results to observational bounds.  We found that a large fraction of models are allowed by current constraints.

We conclude that in the absence of vectorlike pairs, the strong CP problem is generically solved in this landscape of theories.\\

\section*{Acknowledgements}

We thank Csaba Cs\'{a}ki, Jim Halverson, Doddy Marsh, Matt Reece, Max Ruhrdorfer, and Edward Witten for discussions.  We are grateful to Richard Nally for comments on a draft of this work.
The research of M.D. was supported in part by the National Science Foundation under Cooperative Agreement PHY-2019786 (The NSF AI Institute for Artificial Intelligence and Fundamental Interactions).
The research of M.D., N.G., and L.M.~was supported in part by NSF grant PHY-1719877,
that of C.L. was partially supported by DOE Grant DE-SC0013607, and that of L.M.~and J.M.~was supported in part by the Simons Foundation Origins of the Universe Initiative.

\vfill

\appendix

\section{Computation of masses and decay constants} \label{app:masses_fs}
The Lagrangians considered in this paper take the form
\begin{align}
\mathcal{L} \supset -\frac{1}{2} K_{AB} \partial \xi^A  \partial\xi^B - \sum_\alpha \frac{q^\alpha_A  \xi^A}{32\pi^2} \tilde{G}_\alpha^{\mu \nu} G_{\alpha \mu \nu}- V(\xi^A)
\end{align}
where $V(\xi^A)$ is the scalar potential induced by QCD and stringy instantons, and $G$ denotes non-Abelian gauge field strengths (in particular, QCD). Canonically normalizing the axions $\xi^A$ involves (1) defining a new set of axions $\phi^A={e^A}_B \xi^B$ using a Cholesky decomposition $K=e^{T}\cdot e$ in terms of which the kinetic term is canonical, and (2) using up the residual freedom to rotate the $\phi^A$ with orthogonal transformations in order to diagonalize the Hessian that arises from $V(\xi^A)$.
The first task is computationally straightforward, as the eigenvalues of $K$ turn out not to be hierarchical in our models. The second step, however, is computationally expensive using brute force methods: the non-perturbative scales $\Lambda_{\vec{q}}^4$ entering $V(\xi^A)$ typically vary over many orders of magnitude, so that a direct numerical solution of the eigenvalue problem of the Hessian $\mathcal{H}_{AB}:=\partial_{\xi^A}\partial_{\xi^B}V$ would require working with very high precision.

Instead, we take advantage of the fact that the non-perturbative scales are hierarchical.  Consider a set of $N$ instantons with charges $\vec{q}_A$, $A=1,\ldots,N$, ordered such that $S_E^{\vec{q}_A}<S_E^{\vec{q}_B}$ for all $A<B$. Without loss we assume that the kinetic matrix is the identity (in such a basis the charges $\vec{q}_A$ take real, not integer values). The largest eigenvector of the Hessian clearly receives its dominant contribution from the instanton $\vec{q}_1$ with smallest action, and the largest eigenvector $\delta \phi^A$ is obtained by minimizing the distance in field space required to perturb the instanton phase $\delta\phi \cdot \vec{q}_1$ by a fixed amount, i.e.~$\delta \vec{\phi}=\varphi \frac{\vec{q}_1}{||\vec{q}_1||}$. Along this direction in field space, the axion kinetic term and potential read
\begin{equation}
	\mathcal{L}\supset -\frac{1}{2}(\partial\varphi)^2-\Lambda_{\vec{q}_1}^4\Bigl(1-\cos\bigl(2\pi||\vec{q}_1||\varphi\bigr)\Bigr)\, ,
\end{equation}
so one finds a decay constant\footnote{The notion of a ``decay constant" is not well-defined for an axion potential with more than $N$ instanton contributions, but the quantities $f_i$ defined in this Appendix are nonetheless good proxies for the axion field range when the masses are exponentially hierarchical, as they are here. As such, we will refer to these quantities as ``decay constants" throughout.} $f_1=(2\pi||\vec{q}_1||)^{-1}$, and mass $m_1^2=\Lambda_{\vec{q}_1}^4/f_1^2$. Below the mass scale $m_1$ one may integrate out the heaviest axion $\phi$, and parameterize the remaining $N-1$ axion field space as $\delta\phi^A={M^A}_i \delta\hat{\phi}^i$ in terms of $N-1$ low energy axions $\delta\hat{\phi}^i$ where the rows of ${M^A}_i$ generate the kernel of the map $\vec{\phi}\mapsto \vec{q}_1\cdot \vec{\phi}$. Again without loss one may choose $M$ such that $M^T\cdot M=\mathbb{I}_{N-1,N-1}$, and this choice is unique up to an $O(N-1)$ transformation.  The transpose $M^T$ defines a projection $\pi: \, \mathbb{R}^N\rightarrow \mathbb{R}^{N-1}$, $\vec{q}\mapsto M^T\cdot \vec{q}$, and the images of the charges $\vec{q}_{2,\ldots,N}$ under $\pi$ are the effective instanton charges in the low energy theory. We can repeat the same steps in the low energy theory to determine the next pair of decay constant and mass: $f_2=(2\pi||\pi(\vec{q}_2)||)^{-1}$ and $m_2^2=\Lambda_{\vec{q}_2}^4/f_2^2$. By iterating this all the way to the smallest eigenvalue one finds that the $i$-th decay constant $f_i$ is given by
\begin{equation}
	f_i=(2\pi||\pi_i(\vec{q}_i)||)^{-1}\, ,
\end{equation}
with projections $\pi_i: \, \mathbb{R}^N\rightarrow \mathbb{R}^{N-i+1}$, $\vec{q}\mapsto (M_i)^T\cdot \vec{q}$ and for $i>1$ the rows of the matrices $M_i$ are generators of the kernel of $(\vec{q}_1,\ldots,\vec{q}_{i-1})^T$, chosen such that $(M_i)^T\cdot M_i=\mathbb{I}_{N-i+1,N-i+i}$, while $M_1:=\mathbb{I}_{N,N}$. The masses are then given by $m_i^2=\Lambda_{\vec{q}_i}^4/f_i^2$.

The above analysis gives the correct mass spectrum in the limit $\Lambda_{\vec{q}_A}/\Lambda_{\vec{q}_{A+1}}\rightarrow \infty$. Thus, one expects $\mathcal{O}(1)$ corrections when a pair of instanton actions $S_E^{\vec{q}_A}$, $S_E^{\vec{q}_{A+1}}$ are close to each other, with $S_E^{\vec{q}_{A+1}}-S_E^{\vec{q}_A}\lesssim 1$. Indeed, in our examples this is frequently the case, but in a random sample of $100$ geometries with $h^{1,1}=10$ to $100$, we found deviations in the masses to be $1\%$ on average, with a maximum error of $80\%$, so we deem the above algorithm robust for our purposes.

\section{Sub-leading instantons}\label{app:subleading}

In this Appendix, we explain why the contributions to the scalar potential studied in the main text are indeed the leading ones for the purpose of assessing the Peccei-Quinn quality problem.

\subsubsection{Autochthonous divisors}

The effective divisors on Calabi-Yau threefold hypersurfaces can be categorized as \emph{inherited} or \emph{autochthonous}.
The inherited divisors, of which there are $h^{1,1}+4$, are by definition those that descend from effective divisors of the ambient toric variety.
Autochthonous divisors, on the other hand, are those that do not arise from intersections of the hypersurface with effective divisors of the ambient variety.

Enumerating the inherited divisors in a given geometry is straightforward,
and in the main analysis of \S\ref{sec:quality} we studied the contributions from  Euclidean D3-branes wrapping inherited divisors.
One might then ask whether  Euclidean D3-branes wrapping autochthonous divisors could introduce a PQ quality problem in cases where
Euclidean D3-branes wrapping inherited divisors do not.

Although finding all the autochthonous divisors in a general Calabi-Yau hypersurface is an open problem, a certain type of autochthonous divisor associated to factorizations of the hypersurface defining equation is easily identified from polytope data. We studied the volumes of these so-called `min-face factorized' divisors for every favorable Calabi-Yau hypersurface with $2 \le h^{1,1} \le 4$,
evaluating the divisor volumes at the point in K\"ahler moduli space where all curves have volumes $\ge 1$.
In this test set, we found that the smallest autochthonous divisor is on average $68$ times larger than the smallest inherited divisor, and the smallest such ratio is $4.6$.
Min-face factorized autochthonous divisors were also studied in the compactifications of \cite{Demirtas:2021nlu}, where $51 \le h^{1,1} \le 214$, and the smallest autochthonous divisors was never less than $100$ times larger than the smallest inherited divisor.
Thus, barring the existence of a new class of autochthonous divisors with much smaller volumes, we expect that the PQ quality problem will be dominated by Euclidean D3-branes wrapping inherited divisors.

\subsubsection{K\"ahler potential instantons}\label{kpi}

Thus far we have explained how we compute an upper bound on the effects of the leading \emph{superpotential} terms.
We should still consider whether PQ quality can be affected by instanton corrections to the K\"ahler potential that arise from Euclidean D3-branes wrapping \textit{non-holomorphic} 4-cycles in the Calabi-Yau.

Suppose that $[\Sigma] \in H_4(X, \mathbb{Z})$ is a homology class that admits no holomorphic representative.
Then a Euclidean D3-brane in the class $[\Sigma]$ will wrap the locally volume-minimizing representative $\Sigma_{\text{min}}$ of $[\Sigma]$.
Because $\Sigma_{\text{min}}$ is not calibrated by the K\"ahler form, computing its volume is difficult.
However, any class $[\Sigma]$ can be represented as a sum of classes that admit holomorphic or anti-holomorphic representatives, and so $[\Sigma]$ has
a representative  $\Sigma_{\cup}$ that is a union of holomorphic and anti-holomorphic cycles.  We call such a representative a \textit{piecewise calibrated} representative.
The piecewise-calibrated representative is generally \emph{not} locally volume-minimizing, but one
can think of the minimum-volume representative in that class as a fusing, or recombination, of the components of the piecewise-calibrated representative.  We define the recombination fraction \cite{Demirtas:2019lfi}
\begin{equation}\label{rdef}
\mathfrak{r}_{\Sigma} := \frac{\text{Vol}(\Sigma_\cup) - \text{Vol}(\Sigma_{\text{min}})}{\text{Vol}(\Sigma_{\text{min}})}\,.
\end{equation}

We would like to know whether a  contribution to the K\"ahler potential from a Euclidean D3-brane wrapping a non-holomorphic cycle $\Sigma_{\text{min}}$ in some class $[\Sigma]$ can have a larger impact on PQ quality than that of any possible superpotential term in the same compactification.  This can occur only when the recombination \eqref{rdef} is significant, $\mathfrak{r}_{\Sigma} \gtrsim 1$, as otherwise the superpotential terms corresponding to the holomorphic and antiholomorphic components of $\Sigma_{\cup}$ --- which were already incorporated in the analysis of \S\ref{sec:quality} --- will have effects that are at least comparable to the K\"ahler potential term.

Thus, very large recombination could spoil the PQ mechanism.  However, there is no evidence for large recombination in the substantial literature on geometric measure theory: see e.g.~\cite{Micallef2006}.  Moreover, the investigations of \cite{Demirtas:2019lfi, Long:2021lon}, which aimed in part to identify examples with large recombination, were able to show that (rather strong forms of) the Weak Gravity Conjecture (WGC) can imply \emph{some} degree of recombination, but not enough to invalidate our conclusions about PQ quality. Indeed, because the axionic form of the WGC generally implies a bound $S\lesssim M_\mathrm{pl}/f$, while the decay constants considered here are far below the Planck scale, it is implausible that WGC constraints alone could be stringent enough to impact PQ quality.  Finally, one should recognize that if the effects of K\"ahler potential instantons are generally parametrically enhanced by large recombination --- a conclusion that we judge to be very unlikely on the basis of the evidence given above --- then this would invalidate not just our conclusions, but also much of the literature on four-dimensional theories arising in string compactifications.  We leave this intriguing but implausible possibility as a question for the future.

\bibliographystyle{JHEP}
\bibliography{refs}

\providecommand{\href}[2]{#2}\begingroup\raggedright\begin{thebibliography}{10}

\bibitem{nEDM:2020crw}
{\scshape nEDM} collaboration, C.~Abel et~al., \emph{{Measurement of the
  permanent electric dipole moment of the neutron}},
  \href{https://doi.org/10.1103/PhysRevLett.124.081803}{\emph{Phys. Rev. Lett.}
  {\bfseries 124} (2020) 081803},
  [\href{https://arxiv.org/abs/2001.11966}{{\ttfamily 2001.11966}}].

\bibitem{PhysRevLett.97.131801}
C.~A. Baker, D.~D. Doyle, P.~Geltenbort, K.~Green, M.~G.~D. van~der Grinten,
  P.~G. Harris et~al., \emph{Improved experimental limit on the electric dipole
  moment of the neutron},
  \href{https://doi.org/10.1103/PhysRevLett.97.131801}{\emph{Phys. Rev. Lett.}
  {\bfseries 97} (Sep, 2006) 131801}.

\bibitem{PhysRevD.92.092003}
J.~M. Pendlebury, S.~Afach, N.~J. Ayres, C.~A. Baker, G.~Ban, G.~Bison et~al.,
  \emph{Revised experimental upper limit on the electric dipole moment of the
  neutron}, \href{https://doi.org/10.1103/PhysRevD.92.092003}{\emph{Phys. Rev.
  D} {\bfseries 92} (Nov, 2015) 092003}.

\bibitem{Peccei:1977hh}
R.~D. Peccei and H.~R. Quinn, \emph{{CP Conservation in the Presence of
  Instantons}}, \href{https://doi.org/10.1103/PhysRevLett.38.1440}{\emph{Phys.
  Rev. Lett.} {\bfseries 38} (1977) 1440--1443}.

\bibitem{GrillidiCortona:2015jxo}
G.~Grilli~di Cortona, E.~Hardy, J.~Pardo~Vega and G.~Villadoro, \emph{{The QCD
  axion, precisely}},
  \href{https://doi.org/10.1007/JHEP01(2016)034}{\emph{JHEP} {\bfseries 01}
  (2016) 034}, [\href{https://arxiv.org/abs/1511.02867}{{\ttfamily
  1511.02867}}].

\bibitem{Georgi:1981pu}
H.~M. Georgi, L.~J. Hall and M.~B. Wise, \emph{{Grand Unified Models With an
  Automatic {Peccei-Quinn} Symmetry}},
  \href{https://doi.org/10.1016/0550-3213(81)90433-8}{\emph{Nucl. Phys. B}
  {\bfseries 192} (1981) 409--416}.

\bibitem{Holman:1992us}
R.~Holman, S.~D.~H. Hsu, T.~W. Kephart, E.~W. Kolb, R.~Watkins and L.~M.
  Widrow, \emph{{Solutions to the strong CP problem in a world with gravity}},
  \href{https://doi.org/10.1016/0370-2693(92)90491-L}{\emph{Phys. Lett. B}
  {\bfseries 282} (1992) 132--136},
  [\href{https://arxiv.org/abs/hep-ph/9203206}{{\ttfamily hep-ph/9203206}}].

\bibitem{PhysRevD.46.539}
S.~M. Barr and D.~Seckel, \emph{Planck-scale corrections to axion models},
  \href{https://doi.org/10.1103/PhysRevD.46.539}{\emph{Phys. Rev. D} {\bfseries
  46} (Jul, 1992) 539--549}.

\bibitem{Kamionkowski:1992mf}
M.~Kamionkowski and J.~March-Russell, \emph{{Planck scale physics and the
  Peccei-Quinn mechanism}},
  \href{https://doi.org/10.1016/0370-2693(92)90492-M}{\emph{Phys. Lett. B}
  {\bfseries 282} (1992) 137--141},
  [\href{https://arxiv.org/abs/hep-th/9202003}{{\ttfamily hep-th/9202003}}].

\bibitem{Conlon:2006tq}
J.~P. Conlon, \emph{{The QCD axion and moduli stabilisation}},
  \href{https://doi.org/10.1088/1126-6708/2006/05/078}{\emph{JHEP} {\bfseries
  05} (2006) 078}, [\href{https://arxiv.org/abs/hep-th/0602233}{{\ttfamily
  hep-th/0602233}}].

\bibitem{plauschinn}
R.~Blumenhagen, S.~Moster and E.~Plauschinn, \emph{{Moduli Stabilisation versus
  Chirality for MSSM like Type IIB Orientifolds}},
  \href{https://doi.org/10.1088/1126-6708/2008/01/058}{\emph{JHEP} {\bfseries
  01} (2008) 058}, [\href{https://arxiv.org/abs/0711.3389}{{\ttfamily
  0711.3389}}].

\bibitem{Grimm:2011dj}
T.~W. Grimm, M.~Kerstan, E.~Palti and T.~Weigand, \emph{{On Fluxed Instantons
  and Moduli Stabilisation in IIB Orientifolds and F-theory}},
  \href{https://doi.org/10.1103/PhysRevD.84.066001}{\emph{Phys. Rev. D}
  {\bfseries 84} (2011) 066001},
  [\href{https://arxiv.org/abs/1105.3193}{{\ttfamily 1105.3193}}].

\bibitem{Broeckel:2021dpz}
I.~Broeckel, M.~Cicoli, A.~Maharana, K.~Singh and K.~Sinha, \emph{{Moduli
  stabilisation and the statistics of axion physics in the landscape}},
  \href{https://doi.org/10.1007/JHEP08(2021)059}{\emph{JHEP} {\bfseries 08}
  (2021) 059}, [\href{https://arxiv.org/abs/2105.02889}{{\ttfamily
  2105.02889}}].

\bibitem{Bies:2021nje}
M.~Bies, M.~Cveti\v{c}, R.~Donagi, M.~Liu and M.~Ong, \emph{{Root Bundles and
  Towards Exact Matter Spectra of F-theory MSSMs}},
  \href{https://arxiv.org/abs/2102.10115}{{\ttfamily 2102.10115}}.

\bibitem{Preskill:1982cy}
J.~Preskill, M.~B. Wise and F.~Wilczek, \emph{{Cosmology of the Invisible
  Axion}}, \href{https://doi.org/10.1016/0370-2693(83)90637-8}{\emph{Phys.
  Lett. B} {\bfseries 120} (1983) 127--132}.

\bibitem{Abbott:1982af}
L.~F. Abbott and P.~Sikivie, \emph{{A Cosmological Bound on the Invisible
  Axion}}, \href{https://doi.org/10.1016/0370-2693(83)90638-X}{\emph{Phys.
  Lett. B} {\bfseries 120} (1983) 133--136}.

\bibitem{Dine:1982ah}
M.~Dine and W.~Fischler, \emph{{The Not So Harmless Axion}},
  \href{https://doi.org/10.1016/0370-2693(83)90639-1}{\emph{Phys. Lett. B}
  {\bfseries 120} (1983) 137--141}.

\bibitem{Dine:1986bg}
M.~Dine and N.~Seiberg, \emph{{String Theory and the Strong {CP} Problem}},
  \href{https://doi.org/10.1016/0550-3213(86)90043-X}{\emph{Nucl. Phys. B}
  {\bfseries 273} (1986) 109--124}.

\bibitem{Arkani-Hamed:2005zuc}
N.~Arkani-Hamed, S.~Dimopoulos and S.~Kachru, \emph{{Predictive landscapes and
  new physics at a TeV}},
  \href{https://arxiv.org/abs/hep-th/0501082}{{\ttfamily hep-th/0501082}}.

\bibitem{Dvali:2007hz}
G.~Dvali, \emph{{Black Holes and Large N Species Solution to the Hierarchy
  Problem}}, \href{https://doi.org/10.1002/prop.201000009}{\emph{Fortsch.
  Phys.} {\bfseries 58} (2010) 528--536},
  [\href{https://arxiv.org/abs/0706.2050}{{\ttfamily 0706.2050}}].

\bibitem{Dvali:2007wp}
G.~Dvali and M.~Redi, \emph{{Black Hole Bound on the Number of Species and
  Quantum Gravity at LHC}},
  \href{https://doi.org/10.1103/PhysRevD.77.045027}{\emph{Phys. Rev. D}
  {\bfseries 77} (2008) 045027},
  [\href{https://arxiv.org/abs/0710.4344}{{\ttfamily 0710.4344}}].

\bibitem{Demirtas:2018akl}
M.~Demirtas, C.~Long, L.~McAllister and M.~Stillman, \emph{{The Kreuzer-Skarke
  Axiverse}}, \href{https://doi.org/10.1007/JHEP04(2020)138}{\emph{JHEP}
  {\bfseries 04} (2020) 138},
  [\href{https://arxiv.org/abs/1808.01282}{{\ttfamily 1808.01282}}].

\bibitem{Vafa:1983tf}
C.~Vafa and E.~Witten, \emph{{Restrictions on Symmetry Breaking in Vector-Like
  Gauge Theories}},
  \href{https://doi.org/10.1016/0550-3213(84)90230-X}{\emph{Nucl. Phys. B}
  {\bfseries 234} (1984) 173--188}.

\bibitem{Vafa:1984xg}
C.~Vafa and E.~Witten, \emph{{Parity Conservation in QCD}},
  \href{https://doi.org/10.1103/PhysRevLett.53.535}{\emph{Phys. Rev. Lett.}
  {\bfseries 53} (1984) 535}.

\bibitem{Georgi:1986kr}
H.~Georgi and L.~Randall, \emph{{Flavor Conserving CP Violation in Invisible
  Axion Models}},
  \href{https://doi.org/10.1016/0550-3213(86)90022-2}{\emph{Nucl. Phys. B}
  {\bfseries 276} (1986) 241--252}.

\bibitem{Holdom:1985vx}
B.~Holdom, \emph{{Strong QCD at High-energies and a Heavy Axion}},
  \href{https://doi.org/10.1016/0370-2693(85)90371-5}{\emph{Phys. Lett. B}
  {\bfseries 154} (1985) 316}.

\bibitem{Flynn:1987rs}
J.~M. Flynn and L.~Randall, \emph{{A Computation of the Small Instanton
  Contribution to the Axion Potential}},
  \href{https://doi.org/10.1016/0550-3213(87)90089-7}{\emph{Nucl. Phys. B}
  {\bfseries 293} (1987) 731--739}.

\bibitem{Choi:1998ep}
K.~Choi and H.~D. Kim, \emph{{Small instanton contribution to the axion
  potential in supersymmetric models}},
  \href{https://doi.org/10.1103/PhysRevD.59.072001}{\emph{Phys. Rev. D}
  {\bfseries 59} (1999) 072001},
  [\href{https://arxiv.org/abs/hep-ph/9809286}{{\ttfamily hep-ph/9809286}}].

\bibitem{Csaki:2019vte}
C.~Cs\'aki, M.~Ruhdorfer and Y.~Shirman, \emph{{UV Sensitivity of the Axion
  Mass from Instantons in Partially Broken Gauge Groups}},
  \href{https://doi.org/10.1007/JHEP04(2020)031}{\emph{JHEP} {\bfseries 04}
  (2020) 031}, [\href{https://arxiv.org/abs/1912.02197}{{\ttfamily
  1912.02197}}].

\bibitem{Gherghetta:2020keg}
T.~Gherghetta, V.~V. Khoze, A.~Pomarol and Y.~Shirman, \emph{{The Axion Mass
  from 5D Small Instantons}},
  \href{https://doi.org/10.1007/JHEP03(2020)063}{\emph{JHEP} {\bfseries 03}
  (2020) 063}, [\href{https://arxiv.org/abs/2001.05610}{{\ttfamily
  2001.05610}}].

\bibitem{Kitano:2021fdl}
R.~Kitano and W.~Yin, \emph{{Strong CP problem and axion dark matter with small
  instantons}}, \href{https://doi.org/10.1007/JHEP07(2021)078}{\emph{JHEP}
  {\bfseries 07} (2021) 078},
  [\href{https://arxiv.org/abs/2103.08598}{{\ttfamily 2103.08598}}].

\bibitem{Hook:2018dlk}
A.~Hook, \emph{{TASI Lectures on the Strong CP Problem and Axions}},
  {\emph{PoS} {\bfseries TASI2018} (2019) 004},
  [\href{https://arxiv.org/abs/1812.02669}{{\ttfamily 1812.02669}}].

\bibitem{Buchmuller:1985jz}
W.~Buchmuller and D.~Wyler, \emph{{Effective Lagrangian Analysis of New
  Interactions and Flavor Conservation}},
  \href{https://doi.org/10.1016/0550-3213(86)90262-2}{\emph{Nucl. Phys. B}
  {\bfseries 268} (1986) 621--653}.

\bibitem{Canepa:2019hph}
A.~Canepa, \emph{{Searches for Supersymmetry at the Large Hadron Collider}},
  \href{https://doi.org/10.1016/j.revip.2019.100033}{\emph{Rev. Phys.}
  {\bfseries 4} (2019) 100033}.

\bibitem{Cvetic:2011iq}
M.~Cvetic, J.~Halverson and P.~Langacker, \emph{{Implications of String
  Constraints for Exotic Matter and Z' s Beyond the Standard Model}},
  \href{https://doi.org/10.1007/JHEP11(2011)058}{\emph{JHEP} {\bfseries 11}
  (2011) 058}, [\href{https://arxiv.org/abs/1108.5187}{{\ttfamily 1108.5187}}].

\bibitem{Halverson:2013ska}
J.~Halverson, \emph{{Anomaly Nucleation Constrains SU(2) Gauge Theories}},
  \href{https://doi.org/10.1103/PhysRevLett.111.261601}{\emph{Phys. Rev. Lett.}
  {\bfseries 111} (2013) 261601},
  [\href{https://arxiv.org/abs/1310.1091}{{\ttfamily 1310.1091}}].

\bibitem{Svrcek:2006yi}
P.~Svrcek and E.~Witten, \emph{{Axions In String Theory}},
  \href{https://doi.org/10.1088/1126-6708/2006/06/051}{\emph{JHEP} {\bfseries
  06} (2006) 051}, [\href{https://arxiv.org/abs/hep-th/0605206}{{\ttfamily
  hep-th/0605206}}].

\bibitem{Cicoli:2012sz}
M.~Cicoli, M.~Goodsell and A.~Ringwald, \emph{{The type IIB string axiverse and
  its low-energy phenomenology}},
  \href{https://doi.org/10.1007/JHEP10(2012)146}{\emph{JHEP} {\bfseries 10}
  (2012) 146}, [\href{https://arxiv.org/abs/1206.0819}{{\ttfamily 1206.0819}}].

\bibitem{Bachlechner:2019vcb}
T.~C. Bachlechner, K.~Eckerle, O.~Janssen and M.~Kleban, \emph{{The Axidental
  Universe}},  \href{https://arxiv.org/abs/1902.05952}{{\ttfamily 1902.05952}}.

\bibitem{Cicoli:2021gss}
M.~Cicoli, V.~Guidetti, N.~Righi and A.~Westphal, \emph{{Fuzzy Dark Matter
  Candidates from String Theory}},
  \href{https://arxiv.org/abs/2110.02964}{{\ttfamily 2110.02964}}.

\bibitem{Demirtas:2019lfi}
M.~Demirtas, C.~Long, L.~McAllister and M.~Stillman, \emph{{Minimal Surfaces
  and Weak Gravity}},
  \href{https://doi.org/10.1007/JHEP03(2020)021}{\emph{JHEP} {\bfseries 03}
  (2020) 021}, [\href{https://arxiv.org/abs/1906.08262}{{\ttfamily
  1906.08262}}].

\bibitem{Long:2021lon}
C.~Long, A.~Sheshmani, C.~Vafa and S.-T. Yau, \emph{{Non-Holomorphic Cycles and
  Non-BPS Black Branes}},  \href{https://arxiv.org/abs/2104.06420}{{\ttfamily
  2104.06420}}.

\bibitem{Kreuzer:2000xy}
M.~Kreuzer and H.~Skarke, \emph{{Complete classification of reflexive polyhedra
  in four-dimensions}},
  \href{https://doi.org/10.4310/ATMP.2000.v4.n6.a2}{\emph{Adv. Theor. Math.
  Phys.} {\bfseries 4} (2002) 1209--1230},
  [\href{https://arxiv.org/abs/hep-th/0002240}{{\ttfamily hep-th/0002240}}].

\bibitem{Batyrev}
V.~V. Batyrev, \emph{{Dual Polyhedra and Mirror Symmetry for Calabi-Yau
  Hypersurfaces in Toric Varieties}}, {\emph{{J. Alg. Geom}} (1996) 493--535},
  [\href{https://arxiv.org/abs/alg-geom/9310003}{{\ttfamily
  alg-geom/9310003}}].

\bibitem{Demirtas:2020dbm}
M.~Demirtas, L.~McAllister and A.~Rios-Tascon, \emph{{Bounding the
  Kreuzer-Skarke Landscape}},
  \href{https://arxiv.org/abs/2008.01730}{{\ttfamily 2008.01730}}.

\bibitem{CYTools}
M.~Demirtas, L.~McAllister and A.~Rios-Tascon, \emph{{{\tt{CYTools}}: A
  Software Package for Analyzing Calabi-Yau Manifolds, to appear}}.

\bibitem{Halverson:2019cmy}
J.~Halverson, C.~Long, B.~Nelson and G.~Salinas, \emph{{Towards string theory
  expectations for photon couplings to axionlike particles}},
  \href{https://doi.org/10.1103/PhysRevD.100.106010}{\emph{Phys. Rev. D}
  {\bfseries 100} (2019) 106010},
  [\href{https://arxiv.org/abs/1909.05257}{{\ttfamily 1909.05257}}].

\bibitem{Mehta:2020kwu}
V.~M. Mehta, M.~Demirtas, C.~Long, D.~J.~E. Marsh, L.~Mcallister and M.~J.
  Stott, \emph{{Superradiance Exclusions in the Landscape of Type IIB String
  Theory}},  \href{https://arxiv.org/abs/2011.08693}{{\ttfamily 2011.08693}}.

\bibitem{Mehta:2021pwf}
V.~M. Mehta, M.~Demirtas, C.~Long, D.~J.~E. Marsh, L.~McAllister and M.~J.
  Stott, \emph{{Superradiance in string theory}},
  \href{https://doi.org/10.1088/1475-7516/2021/07/033}{\emph{JCAP} {\bfseries
  07} (2021) 033}, [\href{https://arxiv.org/abs/2103.06812}{{\ttfamily
  2103.06812}}].

\bibitem{Veneziano:1982ah}
G.~Veneziano and S.~Yankielowicz, \emph{{An Effective Lagrangian for the Pure
  N=1 Supersymmetric Yang-Mills Theory}},
  \href{https://doi.org/10.1016/0370-2693(82)90828-0}{\emph{Phys. Lett. B}
  {\bfseries 113} (1982) 231}.

\bibitem{Affleck:1983mk}
I.~Affleck, M.~Dine and N.~Seiberg, \emph{{Dynamical Supersymmetry Breaking in
  Supersymmetric QCD}},
  \href{https://doi.org/10.1016/0550-3213(84)90058-0}{\emph{Nucl. Phys. B}
  {\bfseries 241} (1984) 493--534}.

\bibitem{Denef:2005mm}
F.~Denef, M.~R. Douglas, B.~Florea, A.~Grassi and S.~Kachru, \emph{{Fixing all
  moduli in a simple f-theory compactification}},
  \href{https://doi.org/10.4310/ATMP.2005.v9.n6.a1}{\emph{Adv. Theor. Math.
  Phys.} {\bfseries 9} (2005) 861--929},
  [\href{https://arxiv.org/abs/hep-th/0503124}{{\ttfamily hep-th/0503124}}].

\bibitem{Demirtas:2021nlu}
M.~Demirtas, M.~Kim, L.~McAllister, J.~Moritz and A.~Rios-Tascon, \emph{{Small
  Cosmological Constants in String Theory}},
  \href{https://arxiv.org/abs/2107.09064}{{\ttfamily 2107.09064}}.

\bibitem{Marsh:2015xka}
D.~J.~E. Marsh, \emph{{Axion Cosmology}},
  \href{https://doi.org/10.1016/j.physrep.2016.06.005}{\emph{Phys. Rept.}
  {\bfseries 643} (2016) 1--79},
  [\href{https://arxiv.org/abs/1510.07633}{{\ttfamily 1510.07633}}].

\bibitem{Cyncynates:2021yjw}
D.~Cyncynates, T.~Giurgica-Tiron, O.~Simon and J.~O. Thompson,
  \emph{{Friendship in the Axiverse: Late-time direct and astrophysical
  signatures of early-time nonlinear axion dynamics}},
  \href{https://arxiv.org/abs/2109.09755}{{\ttfamily 2109.09755}}.

\bibitem{Chadha-Day:2021uyt}
F.~Chadha-Day, \emph{{Axion-like particle oscillations}},
  \href{https://doi.org/10.1088/1475-7516/2022/01/013}{\emph{JCAP} {\bfseries
  01} (2022) 013}, [\href{https://arxiv.org/abs/2107.12813}{{\ttfamily
  2107.12813}}].

\bibitem{Marsh:2010wq}
D.~J.~E. Marsh and P.~G. Ferreira, \emph{{Ultra-Light Scalar Fields and the
  Growth of Structure in the Universe}},
  \href{https://doi.org/10.1103/PhysRevD.82.103528}{\emph{Phys. Rev. D}
  {\bfseries 82} (2010) 103528},
  [\href{https://arxiv.org/abs/1009.3501}{{\ttfamily 1009.3501}}].

\bibitem{CAST:2017uph}
{\scshape CAST} collaboration, V.~Anastassopoulos et~al., \emph{{New CAST Limit
  on the Axion-Photon Interaction}},
  \href{https://doi.org/10.1038/nphys4109}{\emph{Nature Phys.} {\bfseries 13}
  (2017) 584--590}, [\href{https://arxiv.org/abs/1705.02290}{{\ttfamily
  1705.02290}}].

\bibitem{Wouters:2013hua}
D.~Wouters and P.~Brun, \emph{{Constraints on Axion-like Particles from X-Ray
  Observations of the Hydra Galaxy Cluster}},
  \href{https://doi.org/10.1088/0004-637X/772/1/44}{\emph{Astrophys. J.}
  {\bfseries 772} (2013) 44},
  [\href{https://arxiv.org/abs/1304.0989}{{\ttfamily 1304.0989}}].

\bibitem{Berg:2016ese}
M.~Berg, J.~P. Conlon, F.~Day, N.~Jennings, S.~Krippendorf, A.~J. Powell
  et~al., \emph{{Constraints on Axion-Like Particles from X-ray Observations of
  NGC1275}}, \href{https://doi.org/10.3847/1538-4357/aa8b16}{\emph{Astrophys.
  J.} {\bfseries 847} (2017) 101},
  [\href{https://arxiv.org/abs/1605.01043}{{\ttfamily 1605.01043}}].

\bibitem{Marsh:2017yvc}
M.~C.~D. Marsh, H.~R. Russell, A.~C. Fabian, B.~P. McNamara, P.~Nulsen and
  C.~S. Reynolds, \emph{{A New Bound on Axion-Like Particles}},
  \href{https://doi.org/10.1088/1475-7516/2017/12/036}{\emph{JCAP} {\bfseries
  12} (2017) 036}, [\href{https://arxiv.org/abs/1703.07354}{{\ttfamily
  1703.07354}}].

\bibitem{Conlon:2017qcw}
J.~P. Conlon, F.~Day, N.~Jennings, S.~Krippendorf and M.~Rummel,
  \emph{{Constraints on Axion-Like Particles from Non-Observation of Spectral
  Modulations for X-ray Point Sources}},
  \href{https://doi.org/10.1088/1475-7516/2017/07/005}{\emph{JCAP} {\bfseries
  07} (2017) 005}, [\href{https://arxiv.org/abs/1704.05256}{{\ttfamily
  1704.05256}}].

\bibitem{Chen:2017mjf}
L.~Chen and J.~P. Conlon, \emph{{Constraints on massive axion-like particles
  from X-ray observations of NGC 1275}},
  \href{https://doi.org/10.1093/mnras/sty1591}{\emph{Mon. Not. Roy. Astron.
  Soc.} {\bfseries 479} (2018) 2243--2248},
  [\href{https://arxiv.org/abs/1712.08313}{{\ttfamily 1712.08313}}].

\bibitem{Reynolds:2019uqt}
C.~S. Reynolds, M.~C.~D. Marsh, H.~R. Russell, A.~C. Fabian, R.~Smith,
  F.~Tombesi et~al., \emph{{Astrophysical limits on very light axion-like
  particles from Chandra grating spectroscopy of NGC 1275}},
  \href{https://arxiv.org/abs/1907.05475}{{\ttfamily 1907.05475}}.

\bibitem{Dine:2018glh}
M.~Dine, L.~Stephenson~Haskins, L.~Ubaldi and D.~Xu, \emph{{Some Remarks on
  Anthropic Approaches to the Strong CP Problem}},
  \href{https://doi.org/10.1007/JHEP05(2018)171}{\emph{JHEP} {\bfseries 05}
  (2018) 171}, [\href{https://arxiv.org/abs/1801.03466}{{\ttfamily
  1801.03466}}].

\bibitem{Polchinski:2006gy}
J.~Polchinski, \emph{{The Cosmological Constant and the String Landscape}},  in
  \emph{{23rd Solvay Conference in Physics: The Quantum Structure of Space and
  Time}}, pp.~216--236, 3, 2006,
  \href{https://arxiv.org/abs/hep-th/0603249}{{\ttfamily hep-th/0603249}}.

\bibitem{Micallef2006}
M.~Micallef and J.~Wolfson, \emph{{Area minimizers in a K3 surface and
  holomorphicity}}, \href{https://doi.org/10.1007/s00039-006-0555-x}{\emph{GAFA
  Geometric And Functional Analysis} {\bfseries 16} (01, 2006) 437--452}.

\end{thebibliography}\endgroup

\end{document}